\documentclass[aps,twocolumn,nofootinbib,superscriptaddress]{revtex4-1}

\usepackage[utf8]{inputenc}
\usepackage{epsfig}
\usepackage[breaklinks]{hyperref}
\usepackage{nicefrac}
\usepackage{bm}
\usepackage{dcolumn}
\usepackage{amsmath}
\usepackage{amssymb}
\usepackage{isotope}
\usepackage{xspace}
\usepackage{suffix}

\usepackage{color}

\begin{document}

\title{Scattering amplitudes versus potentials in nuclear effective field theory: \\ Search for a potential compromise.}

\author{Manuel Pavon Valderrama}\email{mpavon@buaa.edu.cn}
\affiliation{School of Physics, \\
  International Research Center for Nuclei and Particles in the Cosmos and \\
Beijing Key Laboratory of Advanced Nuclear Materials and Physics, \\
Beihang University, Beijing 100191, China} 

\date{\today}


\begin{abstract} 
  \rule{0ex}{3ex}
  In effective field theory physical quantities, in particular observables,
  are expressed as a power series in terms of a small expansion parameter.
  For non-perturbative systems, for instance nuclear physics,
  this requires the non-perturbative treatment of at least
  part of the interaction (or the potential, if one is dealing with
  a non-relativistic system), while the rest of the interaction
  is included as perturbations.
  This is not entirely trivial and as a consequence different interpretations
  on how to treat these systems have appeared.
  A practical approach is to expand the effective potential,
  where this potential is later fully iterated in the Schr\"odinger equation
  for obtaining amplitudes and observables. The expectation is that
  this will lead to observables that will have an implicit
  power counting expansion.
  Here I explicitly check whether the amplitudes (when expanded according
  to the counting) are actually following the same power counting
  as the potential.
  It happens that reality does not necessarily conform to expectations and
  the amplitudes will sometimes violate the power counting with
  which the potential has been expanded.
  A more formal approach is to formulate the expansion directly
  in terms of amplitudes and observables,
  which is the original aim of the effective field theory idea.
  Yet this second approach is technically complicated. 
  I explore here the possibility of constructing potentials that when fully
  iterated will make sure that amplitudes are indeed expansible
  in terms of a small expansion parameter.
\end{abstract}

\maketitle

\section{Introduction}

The derivation of nuclear forces from first principles is
the central problem of nuclear physics.
It is also a hard nut to crack and despite many breakthroughs
it has not been solved yet, see Ref.~\cite{Machleidt:2017vls}
for a historical perspective.
Nowadays the expression {\it from first principles} refers to
a derivation of the nuclear force grounded on
quantum chromodynamics (QCD),
the fundamental theory of strong interactions.
But QCD is not analytically solvable at the natural energy scales
of hadronic and nuclear physics.
There are strategies to deal with this problem, of which the two most
prominent ones are lattice QCD~\cite{Beane:2010em} and
effective field theory (EFT)~\cite{Bedaque:2002mn,Epelbaum:2008ga,Machleidt:2011zz}.
Lattice QCD directly solves QCD numerically in a space-time lattice.
EFT in contrast handles QCD in an indirect way by exploiting
the symmetries and degrees of freedom of QCD at low energies.
The most fertile idea within the EFT formalism for strong forces
is that of chiral symmetry~\cite{Bernard:1995dp},
which plays a crucial role
in the description of low energy hadronic processes.

EFTs heavily rely on the idea of {\it separation of scales}:
specific physical phenomena take place at a natural low energy scale $Q$,
which can be distinguished from the high energy scale $M$ at which
a more fundamental description will eventually
take place.
For nuclear and hadronic physics this low energy scale is the pion mass
or the typical momenta of nuclei within a nucleus
($Q \sim 100-200\,{\rm MeV}$), while the high energy scale is the typical
mass of most hadrons ($M \sim 0.5-1.0\,{\rm GeV}$)
which is a consequence of QCD.
The key feature of EFTs is that every physical quantity
can be expressed as an expansion in powers of $Q/M$.
The advantages of the EFT expansion cannot be understated:
in principle one has control on the theoretical accuracy of a calculation.
In practice the realization of this idea will inevitably run into
technical complications, particularly in the case of nuclear physics.
The subject of the present manuscript is how to deal with
a few of these complications.

The fact that physical quantities are expansions is incredibly convenient
for {\it precision era nuclear physics} (a term referring to recent
advances in ab-initio methods~\cite{Soma:2012zd,Lahde:2013uqa,Hagen:2013nca,Carlson:2014vla,Binder:2015mbz,Hebeler:2015hla,Hergert:2015awm}),
as this property enables us to understand
the errors of the calculations~\cite{Perez:2014bua,Furnstahl:2015rha,Griesshammer:2015osb}.
Hence the growing interest in developing EFT-based descriptions
of the nuclear force~\cite{Epelbaum:2014sza,Entem:2017gor,Reinert:2017usi}.
This expectation is however at odds with the current implementation
of nuclear EFT, which is a consequence of the history of the field.
A few decades ago the theoretical understanding of EFTs was perturbative
(see Ref.~\cite{Polchinski:1983gv} for a lucid exposition),
as exemplified by chiral perturbation theory,
the EFT for low energy hadronic processes.
However nuclear physics is non-perturbative, a point that is obvious
from the existence of nuclei.
The most popular implementation of nuclear EFT,
the Weinberg counting~\cite{Weinberg:1990rz,Weinberg:1991um},
can be understood in hindsight as an inspired workaround of
the theoretical problem of formulating an EFT
for a non-perturbative physical system.
Weinberg noticed that while amplitudes in nuclear physics are non-perturbative
-- and hence it is not obvious how to expand them in EFT --, 
the nuclear potential is in contrast calculable
in perturbation theory and amenable to an EFT expansion.
Thus a very sensible solution is to simply expand the nuclear potential
in EFT and then use it in the Schr\"odinger equation
as has always been done in nuclear physics.

The problem with this potential-based formulation is
that strictly speaking is not a genuine EFT:
the natural expectation is that a small correction
in the subleading parts of the potential
will translate into a small correction in the amplitudes,
but this has not been tested except in a few instances
usually involving toy models~\cite{Epelbaum:2009sd}.
Besides power counting, there is a second pillar for EFTs:
renormalization group (RG) invariance (or general cutoff independence).
It can be argued indeed that power counting is merely a consequence of
RG invariance~\cite{Birse:1998dk,Birse:2005um,Valderrama:2016koj}.
Yet a series of works have shown that it is probably not possible to meet
the requirements of renormalization with the purely non-perturbative
treatment of the potential~\cite{PavonValderrama:2005wv,PavonValderrama:2005uj}
(which is similar from a certain point of view to
the conclusions of Ref.~\cite{Epelbaum:2009sd}).
As already conjectured in Ref.~\cite{Nogga:2005hy},
it is now known that the mixture of non-perturbative and perturbative methods
can guarantee observables that have a good expansion and are RG invariant
at the same time~\cite{Long:2007vp,Valderrama:2009ei,Valderrama:2011mv,Long:2011qx,Long:2011xw,Long:2011xw,Long:2012ve,Wu:2018lai} (though this is
contingent on implementation, as recently discussed
in~\cite{Gasparyan:2022isg,Peng:2024aiz,Peng:2025ykg,Yang:2024yqv,PavonValderrama:2025tmp,Thim:2025vhe}).
That is, in principle
these calculations are able to lead exactly to what one expects of
nuclear EFT~\footnote{I note in passing that Ref.~\cite{Epelbaum:2009sd}
  claims that non-perturbative renormalization with hard cutoffs
  will lead to a failure of power counting.
  If unqualified, this statement is incorrect: this failure is only known
  to happen if one iterates the subleading pieces of the potential, while
  the iteration of the leading piece is perfectly safe (with no expansion
  in the potential there is no room for a failure in the expansion of
  the amplitudes). Ref.~\cite{Epelbaum:2009sd} also assumes
  the iteration of the subleading pieces and thus
  its conclusions are not applicable to the calculations of
  Refs.~\cite{Nogga:2005hy,Long:2007vp,Valderrama:2009ei,Valderrama:2011mv,Long:2011qx,Long:2011xw,Long:2011xw,Long:2012ve,Wu:2018lai}.
}.

\begin{figure}[ttt]
\begin{center}
  \includegraphics[height=4.5cm]{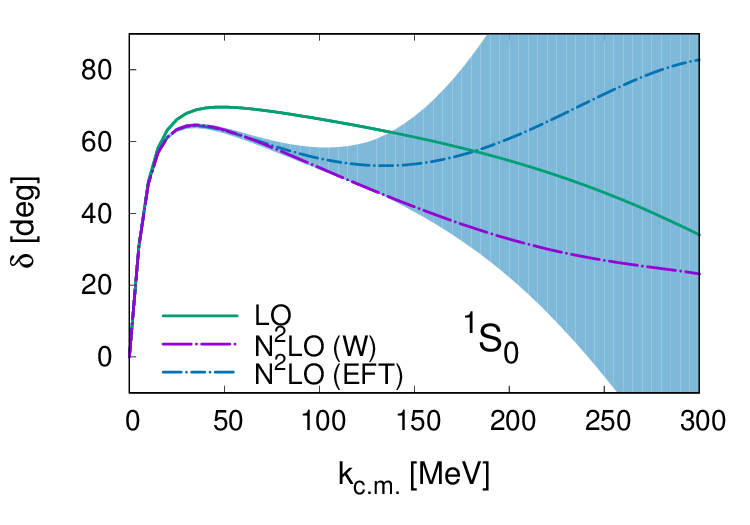}
\end{center}
\caption{Phase shifts in the Weinberg counting as a function of
  the center-of-mass momentum $k_{\rm c.m.}$, depending on whether
  one applies the power counting to the potential (W, for Weinberg
  in the way it is usually applied) or
  to the scattering amplitude (EFT, for indicating that
  the phase shifts are expanded according to the literal EFT expansion
  for the Weinberg counting, Eq.~(\ref{eq:T-exp})).
  The calculation is done at
  ${\rm N^2LO}$ (i.e. including terms up to $\nu=3$ or $Q^3$)
  with a momentum space cutoff $\Lambda = 400\,{\rm MeV}$
  (see Appendix ~\ref{app:p-space}).
  At ${\rm LO}$ the potential- and amplitude-based expansions coincide.
  At ${\rm N^2LO}$ the two expansions are approximately compatible
  within the proposed error bands, which assume a breakdown
  scale of $M = 210\,{\rm MeV}$ (where the details are
  explained around Eqs.~(\ref{eq:err-sum-abridged}) and
  (\ref{eq:err-breakdown}) and in Appendix \ref{app:err}).
}
\label{fig:W1}
\end{figure}

It is worth mentioning that a few authors have conjectured that
the application of non-perturbative methods with a finite
cut-off of an ideal size will lead to a well-defined
expansion~\cite{Epelbaum:2006pt,Epelbaum:2009sd,Epelbaum:2018zli}.
Yet this has only been illustrated with toy models~\cite{Epelbaum:2009sd}.
Needless to say these ideas, though interesting, are not universally
accepted~\cite{Valderrama:2019yiv}, which is why there are still
ongoing efforts for the development of
a fully consistent nuclear EFT (i.e. independent of
regulator choice and with good power counting
features), see Refs.~\cite{Yang:2019hkn,Hammer:2019poc,NavarroPerez:2019sfj,Machleidt:2020vzm,vanKolck:2020llt,Gasparyan:2022isg,Gasparyan:2023rtj,Peng:2024aiz,Yang:2024yqv,Peng:2025ykg} for recent discussions.


\section{EFT expansion}

The purpose of the present manuscript is to check whether
the potential-based approach to nuclear EFT complies
with power counting expectations.
That is, if one is using the Weinberg prescription,
do small corrections to the subleading EFT potential translate
into small corrections to the amplitudes?
In the Weinberg counting~\cite{Weinberg:1990rz,Weinberg:1991um}
the two-nucleon potential is expanded in terms of power counting as
\begin{eqnarray}
  \label{eq:V-exp}
  V = \sum_{\nu = 0}^{\nu_{\rm max}} V^{(\nu)} +
  \mathcal{O}\left( {(\frac{Q}{M})}^{\nu_{\rm max} + 1}) \right) \, ,
\end{eqnarray}
where the specific power counting used is naive dimensional analysis (NDA)
and the error refers to the relative uncertainty.
Depending on how many terms of the expansion one keeps, one will be said to
be working at leading order (${\rm LO}$), next-to-leading order
($\rm NLO$), next-to-next-to-leading order (${\rm N^2LO}$):
\begin{eqnarray}
  V^{\rm LO} &=& V^{(0)} \, , \\
  V^{\rm NLO} &=& V^{(0)} + V^{(2)} \, , \\
  V^{\rm N^2LO} &=& V^{(0)} + V^{(2)} + V^{(3)} \, ,
\end{eqnarray}
plus analogous expressions for higher orders (notice that at $\nu = 1$ there
is no contribution to the EFT potential).
In the Weinberg prescription the scattering amplitude is directly
computed from the EFT potential:
\begin{eqnarray}
  T_{\rm W} = V + V\,G_0\,T_{\rm W} \, ,
\end{eqnarray}
where depending on where one cuts the expansion one will talk about
$T_{\rm W}^{\rm LO}$, $T_{\rm W}^{\rm NLO}$, $T_{\rm W}^{\rm N^2LO}$ and so on.

There are a few technicalities involved in the previous process.
One is that the EFT potential is singular and has to be regularized
for its proper use within the Lippmann-Schwinger equation.
Another is that the EFT potential contains free parameters
that have to be fitted to data.
I will not cover the specific details at the moment
as these are standard technical procedures
(a brief overview is given in Appendices \ref{app:p-space} and \ref{app:r-space}
for p- and r-space, respectively).

\begin{figure}[ttt]
\begin{center}
  \includegraphics[height=4.5cm]{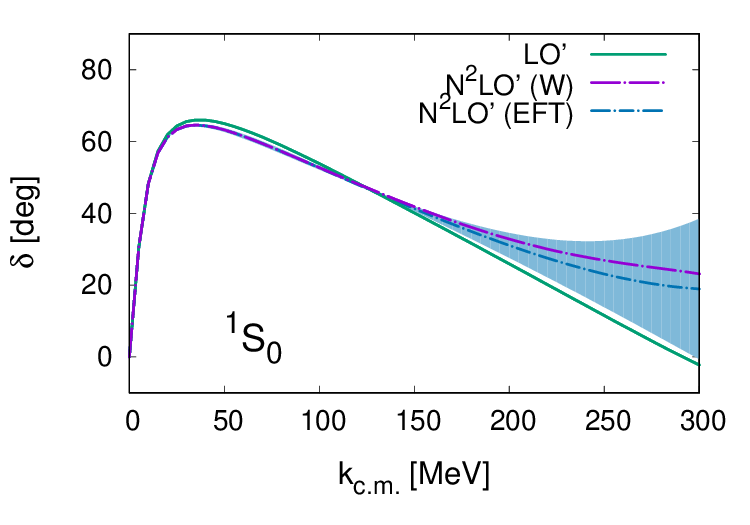}
\end{center}
\caption{Phase shifts in an alternative power counting (read below),
  depending on whether one applies the power counting to the potential (W)
  or to the scattering amplitude (EFT).
  In this alternative power counting I have promoted the leading and
  subleading two pion exchange diagrams to ${\rm LO}$ and demoted
  one-pion exchange to ${\rm N^2LO}$. I use the terms ${\rm LO}'$
  and ${\rm N^2LO'}$ for referring to the orders in this counting.
  This alternative power counting is indistinguishable from the Weinberg one
  when expanding in terms of the potential.
  The error bands of the ${\rm N^2LO' (T)}$ phase shifts assume a breakdown
  scale of $M = 400\,{\rm MeV}$ (see Appendix \ref{app:err}).
}
\label{fig:extravaganza}
\end{figure}

If power counting is really working for the potential, the natural expectation
is that the subleading contributions will be smaller than
the leading order ones for the momenta at which
the expansion is expected to work.
That is, the scattering amplitude can be expanded in terms of the power counting
\begin{eqnarray}
  \label{eq:T-exp}
  T_{\rm EFT} = \sum_{\nu = 0}^{\nu_{\rm max}} T^{(\nu)} +
  \mathcal{O}\left( {(\frac{Q}{M})}^{\nu_{\rm max} + 1} \right) \, ,
\end{eqnarray}
in analogy to what happens with the potential.
Here the EFT label merely refers to the fact that one is expanding,
not to whether the resulting T-matrix is cutoff independent or not
(which depends on the consistency of the power counting
used to expand in the first place).
In turn, expanding means that the subleading contributions to the
scattering amplitude should be perturbative,
where each term in the expansion can be computed as
\begin{eqnarray}
  \label{eq:T0}
  T^{(0)} &=& V^{(0)} + V^{(0)}\,G_0\,T^{(0)} \, , \\
  \label{eq:T2}
  T^{(2)} &=& \left( 1 + T^{(0)}\,G_0 \right)
  \, V^{(2)} \left( \, G_0\,T^{(0)} + 1 \right) , \\
  \label{eq:T3}
  T^{(3)} &=& \left( 1 + T^{(0)}\,G_0 \right)
  \, V^{(3)} \left( \, G_0\,T^{(0)} + 1 \right) , 
\end{eqnarray}
plus more involved expression at higher orders (notice that the expressions
above are relatively simple because the $\nu = 1$ contribution to the EFT
potential cancels, which means that up to $\nu = 3$ first order
perturbation theory is enough).

This means in particular that the {\it a priori} assumptions about
power counting can indeed be tested {\it a posteriori}.
For this, one simply has to compare the T-matrix obtained with the
standard Weinberg prescription with the T-matrix obtained
from the perturbative expansion and check that
their difference is higher order
\begin{eqnarray}
  | T_W - T_{\rm EFT} | \sim
  \mathcal{O}\left( {(\frac{Q}{M})}^{\nu_{\rm max} + 1}) \right) \, .
\end{eqnarray}
Here one only has to adapt the existing frameworks for perturbative
calculations~\cite{Valderrama:2009ei,Valderrama:2011mv,Long:2011qx,Long:2011xw,Long:2012ve}
to this particular case.
In practice, instead of the $T$-matrix, I will compute the phase shift
for a particular partial wave, where for illustrative purposes
I will consider the $^1S_0$ channel at ${\rm N^2LO}$.
The Weinberg amplitude $T_W$ is constructed by fitting the contact-range
couplings of the EFT potential to the $^1S_0$ phase shifts obtained
from a partial wave analysis (PWA) or a phenomenological potential.
In particular I use the Nijmegen II potential~\cite{Stoks:1994wp} ---
equivalent to the Nijmegen PWA~\cite{Stoks:1993tb} --- for this purpose.
The Nijmegen PWA~\cite{Stoks:1993tb} has been recently improved upon
by the more recent PWA of the Granada group~\cite{Perez:2013mwa}
and its phase-equivalent nuclear potentials~\cite{Perez:2013jpa,Perez:2014bua},
but the specific choice of what data to fit is unessential for the discussion.
The ${\rm LO}$ EFT amplitude $T_{\rm EFT}$ is constructed as the Weinberg one
at the same order, while for the subleading corrections I determine
the contact-range couplings by fitting to the phase shifts obtained
from the subleading Weinberg amplitude $T_W$.
That is, at the end I will be comparing theory with theory (in the spirit
of another recent proposal for analyzing power
counting~\cite{Griesshammer:2015osb,Griesshammer:2020fwr}).

The outcome of this analysis can be seen in Fig.~\ref{fig:W1}
for the particular case of the $^{1}S_0$ partial wave,
where a clear breakdown of the assumed power counting
happens at relatively low momenta.
The potential- and amplitude-based expansions use a Gaussian regulator
in p-space with a cutoff $\Lambda = 400\,{\rm MeV}$ (further
details about the calculation are given in the next two paragraphs).
It can be seen that when amplitudes are expanded according to power counting,
the Weinberg power counting is distinctly preserved up to center-of-mass
momenta of $100\,{\rm MeV}$ give or take, and then begins to fail
as the momentum increases.

For a more concrete answer of the momentum at which the breakdown
of the power counting happens, one has to rely on
the uncertainties of the perturbative amplitudes.
Here I define the truncation error of the phase shifts
as the maximum of two sources
\begin{eqnarray}
  \Delta\,\delta_{\rm EFT} &=& {\rm max}
  \bigg[ 
    {\left( \frac{Q}{M} \right)}^{\nu_{\rm max} + 1}\,\sin^2{\delta^{(0)}} ,
    {\left( \frac{Q}{M} \right)}\,\delta^{(\nu_{\rm max})}
    \bigg] \, , \nonumber \\
  \label{eq:err-sum-abridged}
\end{eqnarray}
where in this context $Q$ refers to a specific combination of
the light scales $k$ and $m_{\pi}/2$, which I take to be
$Q^{\nu_{\rm max} + 1} = k^{\nu_{\rm max} +1} + (m_{\pi}/2)^{\nu_{\rm max} + 1}$.
This choice is explained in more detail in Appendix \ref{app:err} and
is very similar in spirit (though not identical) to conventions
used elsewhere in the literature
(e.g. in~\cite{Epelbaum:2014efa,Epelbaum:2014sza,Reinert:2017usi}).
With this, I define the breakdown scale of
the perturbative expansion as the minimum value of $M$ such that
\begin{eqnarray}
  | \delta_W - \delta_{\rm EFT} | \leq \Delta\,\delta_{\rm EFT} \, \quad \,
  \mbox{for $k \leq M$.} \label{eq:err-breakdown}
\end{eqnarray}
This gives $M = 210\,{\rm MeV}$ for the $^1S_0$ channel, where this value is
also used to generate the error bars in Fig.~\ref{fig:W1}.
However, this should only be taken as a tentative answer of the momentum
above which the perturbative reexpansion of the Weinberg amplitudes fails:
a complete answer will require calculations at higher orders, which
are left for future explorations.

I have fitted the amplitude-based expansion to the potential-based
phase shifts in the center-of-mass momentum range
$k_{\rm c.m.} = 20-60\,{\rm MeV}$, where extending the range to higher momenta
does not improve the visual appearance of the fits in Fig.~\ref{fig:W1}
(but definitely spoils the agreement at low momenta).
For further details of the calculation,
I refer to Appendix \ref{app:p-space}. 
I do not show the Nijmegen II phase shifts as they are
not important in this context: they are merely used as a proxy for
the calculation of $T_W$, but the focus is the comparison
between $T_W$ and $T_{\rm EFT}$.
Regarding the $\rm NLO$ phase shifts, I do not show them either
as they are qualitatively similar to the $\rm N^2LO$ ones
(besides, I always fit to the ${\rm N^2LO}$ phases).
I do not consider $\rm N^3LO$ and higher orders in the Weinberg counting
either because they require second order perturbation theory
for the finite-range potential.

To end the discussion, I briefly comment on how the previous breakdown
scale is related with the wider problem of renormalizability
in nuclear EFT.
It is important to stress that I have considered the power counting
expansion of the T-matrix, not its cutoff dependence.
The Weinberg counting does not generate well-defined amplitudes
in the infinite cutoff limit~\cite{Nogga:2005hy} (or zero cutoff limit
if one is working in configuration space) and
this is true regardless of whether one expands
the amplitudes or the potential.
In particular, what I call $T_{\rm EFT}$ in this work will only be cutoff
independent if expanded with a power counting derived
from renormalizability.
The Weinberg counting is not renormalizable though and as the cutoff
increases the perceived breakdown scale of $T_{\rm EFT}$ decreases.
That is, if one is using a non-consistent (i.e. non-renormalizable)
power counting with a hard enough cutoff, then the breakdown scale
could be lower than expected, which is exactly
what is happening in Fig.~\ref{fig:W1}.

The difference with the discussion on renormalizability is the range of
cutoffs involved: while the failure of power counting is apparent
at cutoffs that are pretty common in most EFT calculations,
the failure of renormalizability is only visible at
relatively hard cutoffs, hard enough as to be
considered unphysical.
This is one of the reasons why the problems with the renormalizability
of the Weinberg counting have been sometimes dismissed
as an artifact~\cite{Epelbaum:2006pt,Epelbaum:2018zli}.
But if the problem is power counting instead, then the absence of
excessively hard cutoffs makes it harder to argue
that this is a pseudoproblem.


\section{Power counting breakdown}

As a matter of fact one can use perturbation theory to further
analyze the actual power counting behavior of the amplitudes
against the nominal one in the potential.
While the EFT-expanded phase shifts fail at relatively soft momenta,
this does not happen when the expansion is limited to the potential,
indicating a possible inconsistency or mismatch between
the Weinberg prescription and
the power counting expansion of observables.
Yet, the idea of this type of mismatch is not new and
can be traced further back in time to a popular lecture
by Lepage~\cite{Lepage:1997cs}, which contains a few 
observations about the behavior of EFT operators
when the cutoff is too hard.
For interactions that are strong at short-distances the cutoff
dependence of the couplings can be {\it ``highly nonlinear''},
up to the point that {\it ``an $a^4$ operator can act more
  like an $a^2$ operator, and vice versa''},
Lepage notes~\cite{Lepage:1997cs}, where $a^{n}$
refers to $Q^n$ in the notation used here.

Originally the lecture by Lepage~\cite{Lepage:1997cs} dealt with
the behavior of the contact-range operators in a potential
expansion for hard cutoffs, while here I am dealing
with finite-range interactions.
Leading and subleading two pion exchange, which are in principle $Q^2$ and $Q^3$
contributions to the EFT potential, are known for their unexpected
strength at short distances.
Adapting the observation made by Lepage to finite-range operators,
i.e. the potential, if the cutoff is hard enough it should come
as no surprise that the aforementioned $Q^2$ and $Q^3$ contributions
end up behaving as $Q^0$ operators instead.

For testing this hypothesis I will play the following game: exchange
the orders of one and two pion exchange in the EFT expansion.
That is, I consider a hypothetical power counting in which the
two pion exchange piece of the EFT potential is ${\rm LO}$,
while the one pion exchange piece is demoted to subleading
orders ($\rm N^2LO$ for instance).
The power counting of the contact-range interactions is left unchanged though.
Notice that if one considers only the ${\rm N^2LO}$ potential-based expansion
(i.e. the Weinberg prescription), this hypothetical power counting
is indistinguishable from the Weinberg counting.
The reason is that this calculation only depends on the sum of all
contributions to the potential up to a certain order, but not
on internal rearrangements of these contributions.

The outcome of this game is shown in Fig.~\ref{fig:extravaganza},
which reproduces relatively well the ${\rm N^2LO}$ Weinberg
phase shifts for a breakdown scale of $M = 400\,{\rm MeV}$,
which almost doubles the convergence radius of
the perturbative reexpansion of Fig.~\ref{fig:W1}.
This implies that the actual power counting of the amplitudes is different
than the power counting assumed for the potential, but nonetheless
aligns with the problem described by Lepage in his lectures.
What is surprising though is that the cutoff for which this happens
in the two-nucleon system is $\Lambda = 400\,{\rm MeV}$,
definitely softer than expected.
I will call this type of behavior {\it power counting breakdown}, to
indicate that the amplitudes are breaking the power counting
they are expected to follow in the first place.

The failure of the Weinberg counting (when the amplitudes are
generated by the iteration of the full potential)
could be blamed to a series of causes, of which a few come readily to mind,
such as:
(i) this is a fine-tuned system, (ii) range corrections should
be added when the counting fails, (iii) the one-pion exchange
potential in the singlet vanishes in the chiral limit, etc.
As a matter of fact (i) and (ii) are already an implicit acceptance that
the Weinberg counting might not be applicable in the case of
the $^1S_0$ partial wave, while (iii) might fall
in the category of why this failure happens.
The point is that if one modifies the original Weinberg counting in
line with the different findings derived
from RG analysis~\cite{Nogga:2005hy,Birse:2005um,Valderrama:2009ei,Valderrama:2011mv,Long:2011qx,Long:2011xw,Long:2012ve},
this failure does not happen (as I will explain later).

Yet, it is worth noticing that a different interpretation
is also possible: recently, it has been proposed a reordering of
the nuclear EFT expansion with two-pion exchange
as a leading order interaction~\cite{Mishra:2021luw,Valderrama:2021bql}.
Indeed, ten years prior, this scenario was already considered and calculated
in Ref.~\cite{Valderrama:2010aw}, though as a pathology (not as
the starting point of an alternative power counting).
If this were to be the correct expansion, it could be argued that
the iteration of the full potential simply selects the underlying
power counting of the amplitudes automatically.
  
\begin{figure*}[ttt]
\begin{center}
  \includegraphics[height=4.5cm]{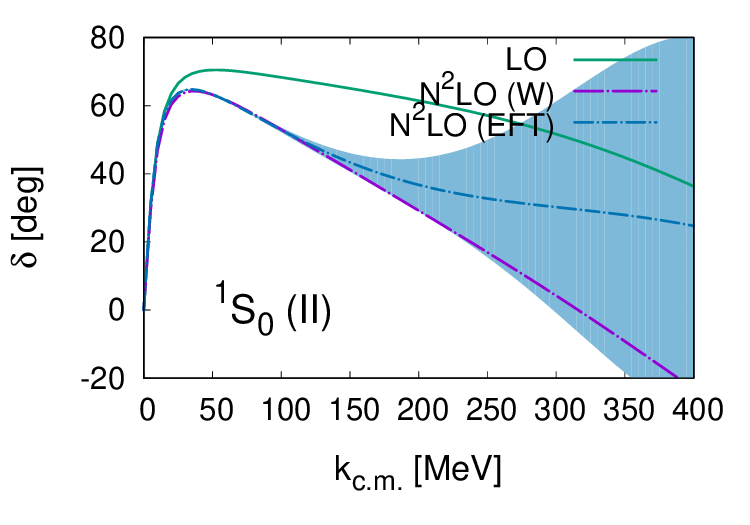}
  \includegraphics[height=4.5cm]{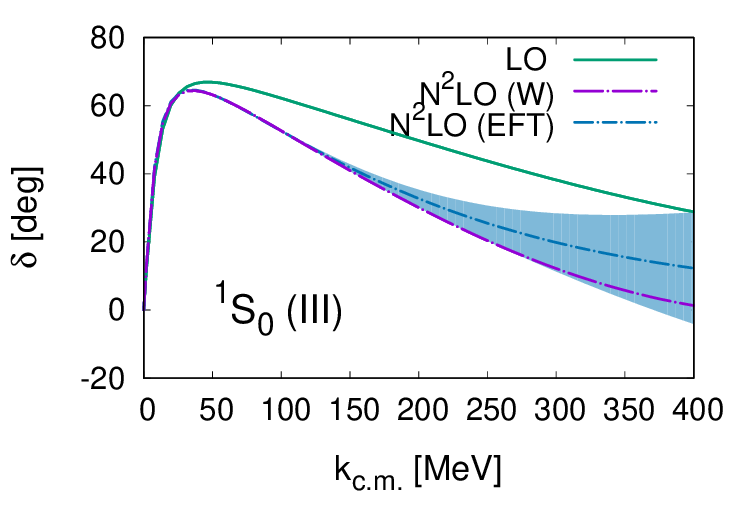}
  \includegraphics[height=4.5cm]{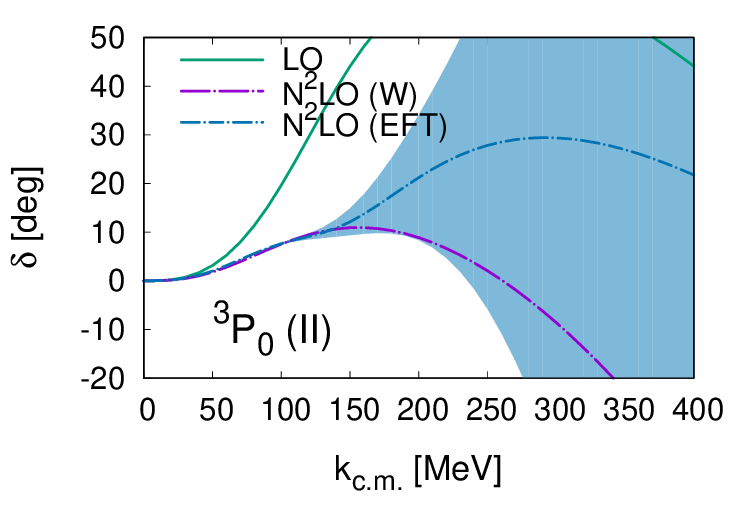}
  \includegraphics[height=4.5cm]{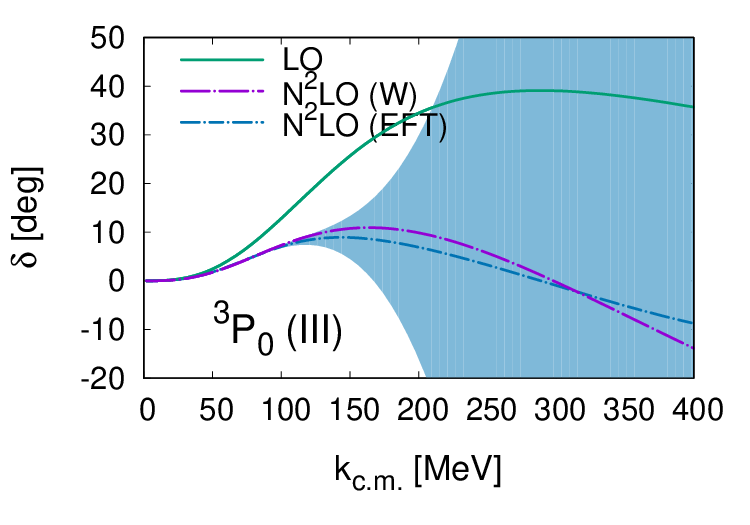}
\end{center}
\caption{Phase shifts for {\it second}~\cite{Epelbaum:2003gr,Epelbaum:2003xx}
  and {\it third generation}~\cite{Gezerlis:2014zia}
  $\rm N^2LO$ potentials in the Weinberg counting, depending
  on whether one expands the potential (W) or the scattering amplitude (EFT).
  For the second generation potential
  of Refs.~\cite{Epelbaum:2003gr,Epelbaum:2003xx}
  the Lippmann-Schwinger and SFR cutoffs fall within
  the $\Lambda = (450-650)$ and
  $\tilde{\Lambda} = (500-700)\,{\rm GeV}$ ranges, where here I use
  $(\Lambda, \tilde{\Lambda}) = (550,600)\,{\rm MeV}$ 
  for obtaining the phase shifts.
  The third generation potential of Ref.~\cite{Gezerlis:2014zia} is formulated
  in coordinate space and uses a Gaussian type cutoff in the range
  $R_c = (1.0-1.2)\,{\rm fm}$ and the SFR cutoff
  $\tilde{\Lambda} = (1.0-1.4)\,{\rm GeV}$.
  In the figures for the potential of Ref.~\cite{Gezerlis:2014zia}
  I have generated the phase shifts for $R_c = 1.1\, {\rm fm}$
  and $\tilde{\Lambda} = 1.0\,{\rm GeV}$.
  I compute the phase shifts for the $^1S_0$ and $^3P_0$ partial waves,
  where the roman numeral in parentheses indicates
  whether the calculation  corresponds to the second (II) or
  third generation (III) potential.
  The errors bands in the amplitude expansion are generated from
  Eqs.~(\ref{eq:err-sum-abridged}) and (\ref{eq:err-breakdown})
  with a breakdown scale of $M = 310\,{\rm MeV}$ and $270\,{\rm MeV}$
  ($M = 380\,{\rm MeV}$ and $190\,{\rm MeV}$) for the $^1S_0$ and
  $^3P_0$ partial waves for the second (third) generation
  potential.
}
\label{fig:W23}
\end{figure*}

\begin{figure*}[ttt]
\begin{center}
  \includegraphics[height=4.5cm]{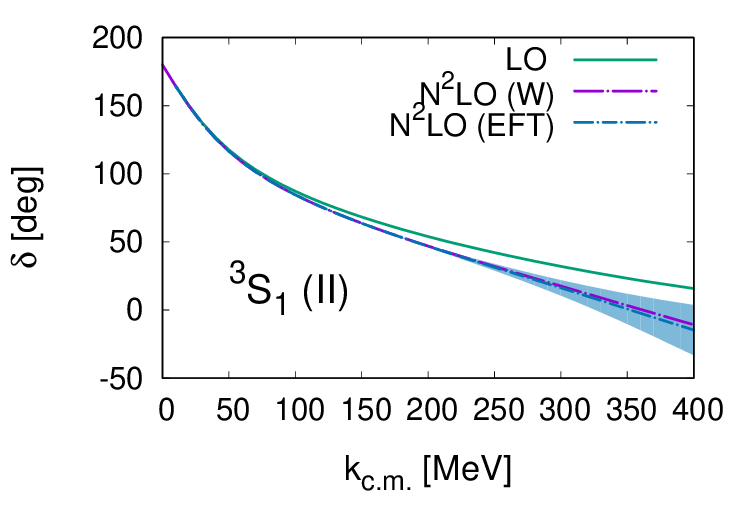}
  \includegraphics[height=4.5cm]{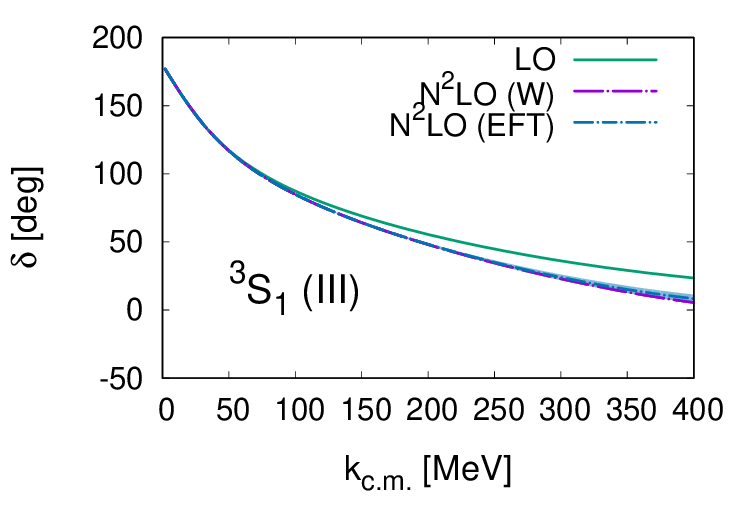}
  \includegraphics[height=4.5cm]{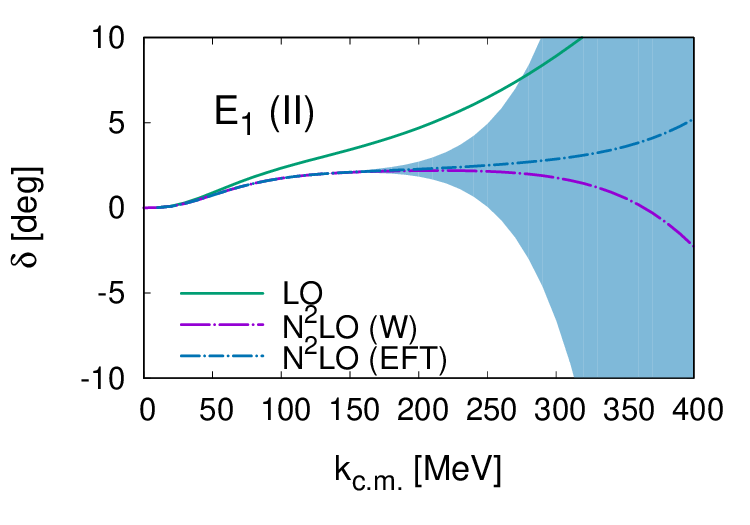}
  \includegraphics[height=4.5cm]{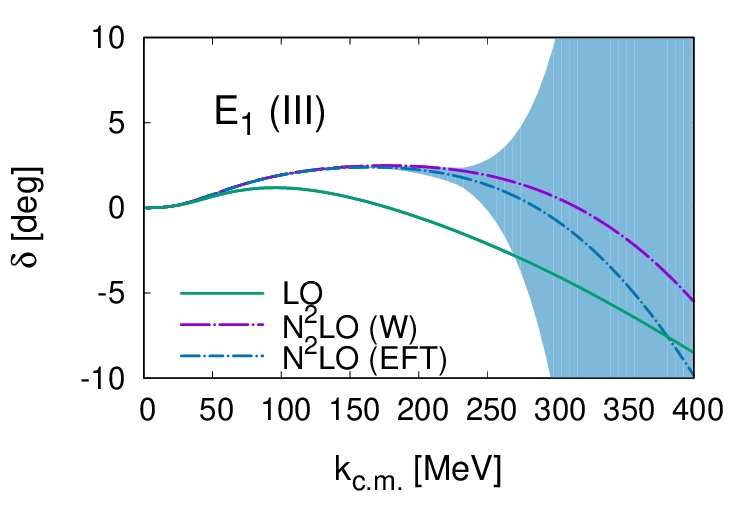}
  \includegraphics[height=4.5cm]{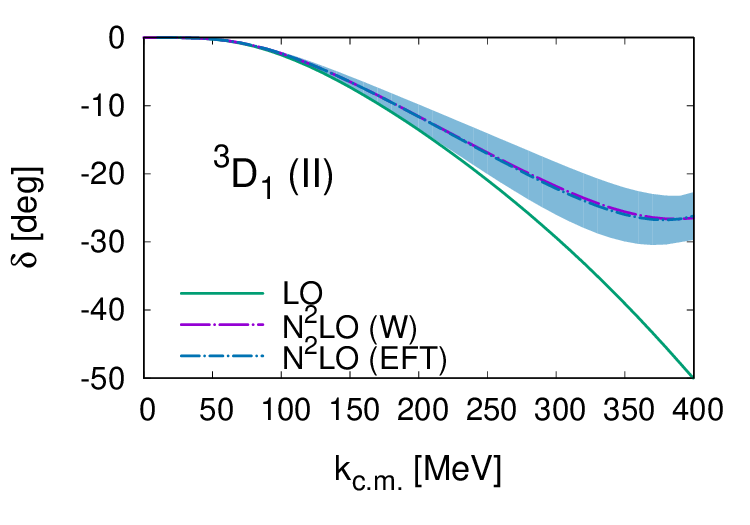}
  \includegraphics[height=4.5cm]{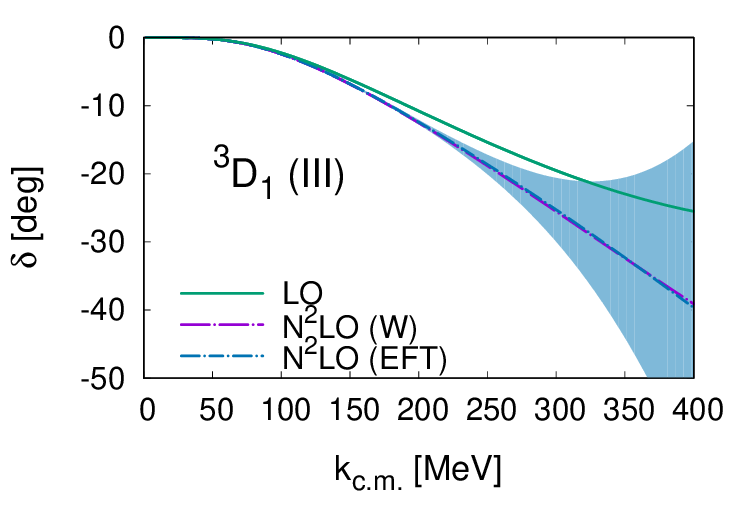}
\end{center}
\caption{Same as Fig.~\ref{fig:W23} but for the ${}^3S_1$, $E_1$ and ${}^3D_1$
   partial waves (nuclear bar phase shifts).
}
\label{fig:W23-3C1}
\end{figure*}

\begin{figure*}[ttt]
\begin{center}
  \includegraphics[height=4.5cm]{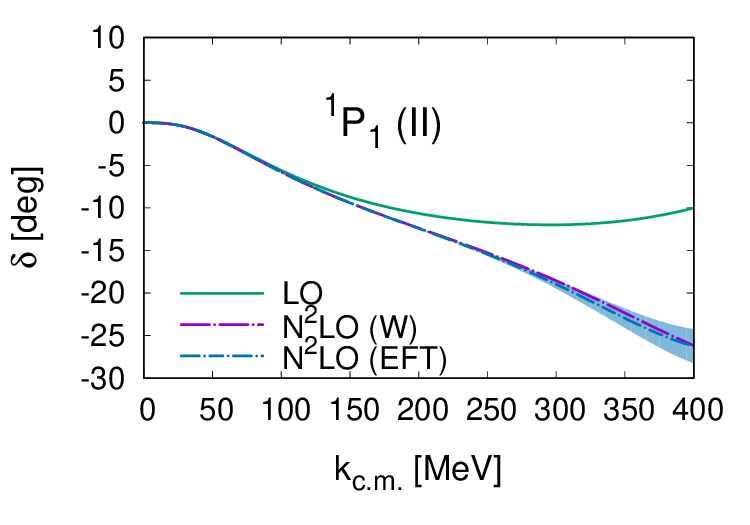}
  \includegraphics[height=4.5cm]{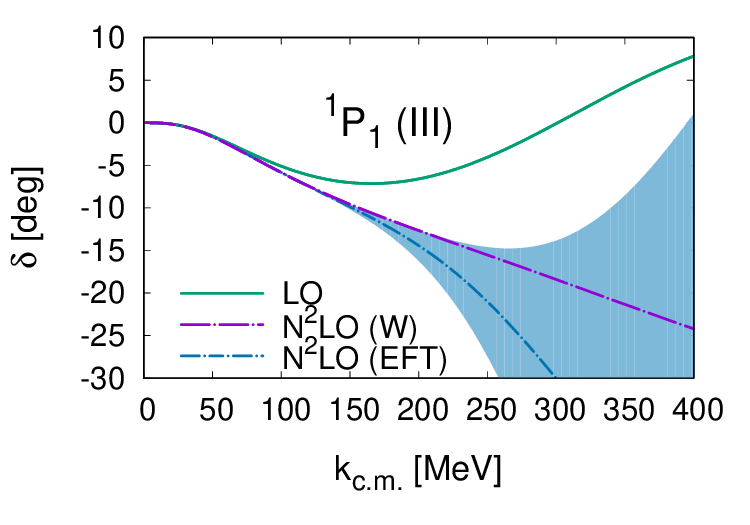}
  \includegraphics[height=4.5cm]{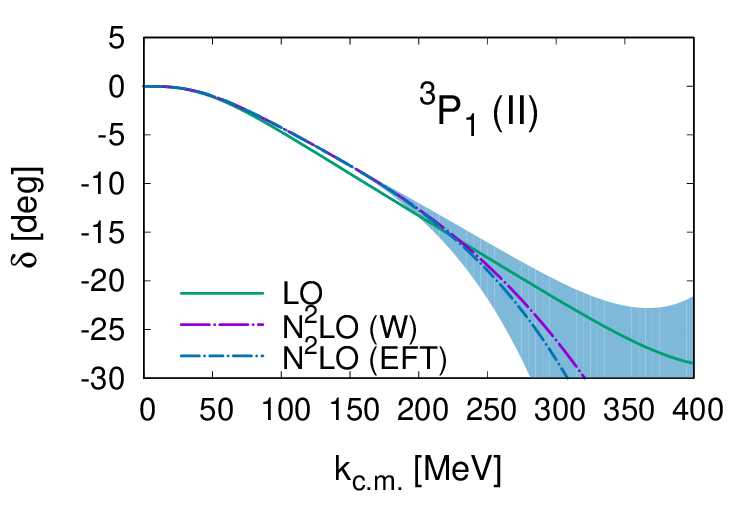}
  \includegraphics[height=4.5cm]{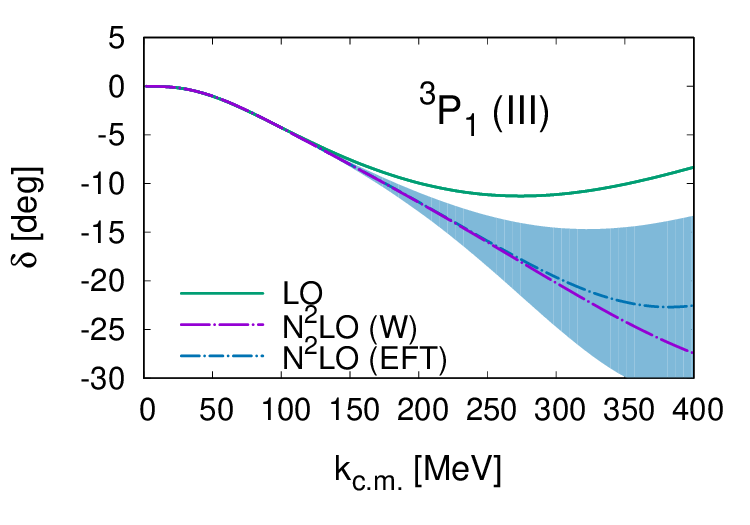}
\end{center}
\caption{Same as Fig.~\ref{fig:W23} but for the ${}^1P_1$, ${}^3P_1$ 
  partial waves.
}
\label{fig:W23-3P}
\end{figure*}

\begin{figure*}[ttt]
\begin{center}
  \includegraphics[height=4.5cm]{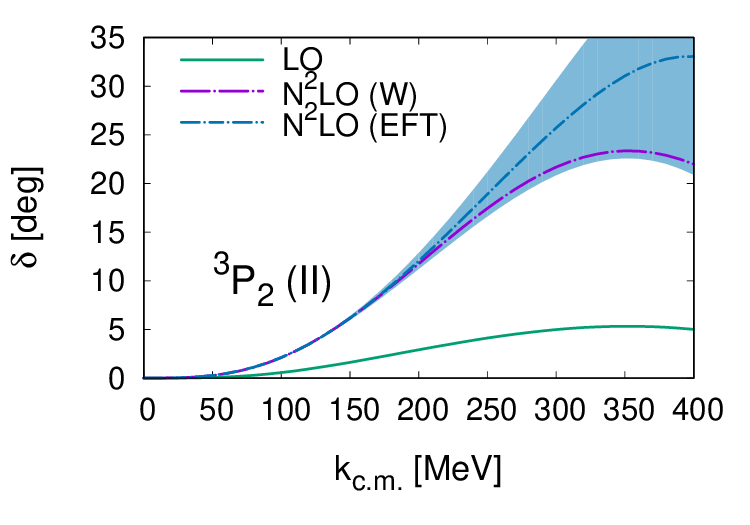}
  \includegraphics[height=4.5cm]{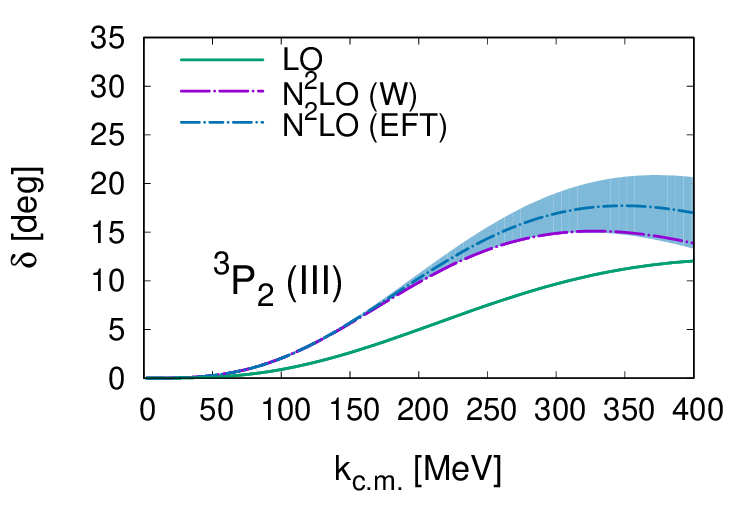}
  \includegraphics[height=4.5cm]{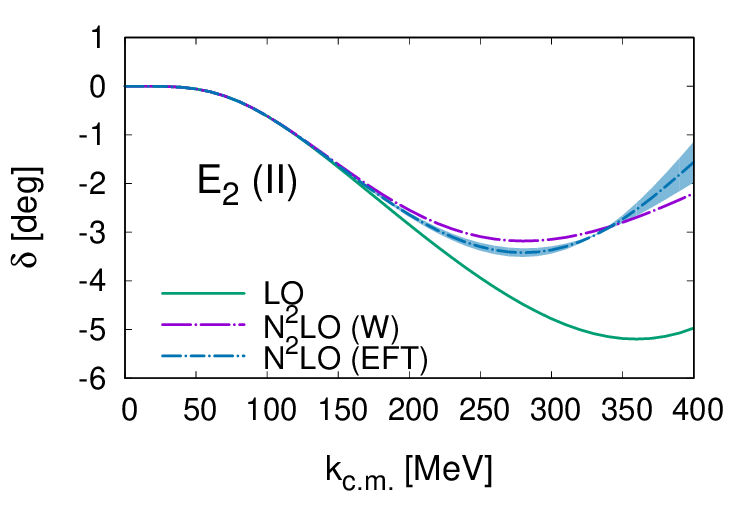}
  \includegraphics[height=4.5cm]{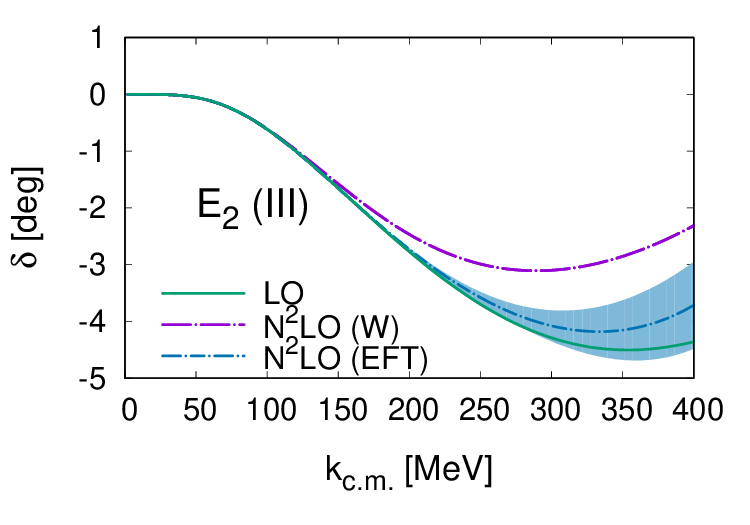}
  \includegraphics[height=4.5cm]{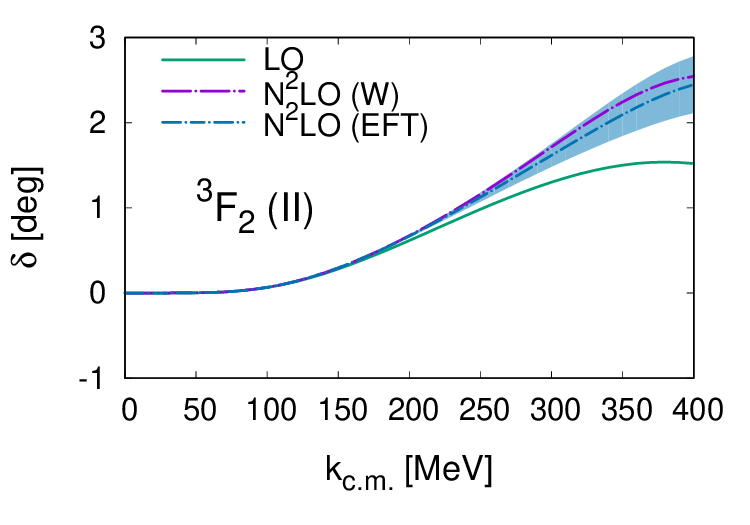}
  \includegraphics[height=4.5cm]{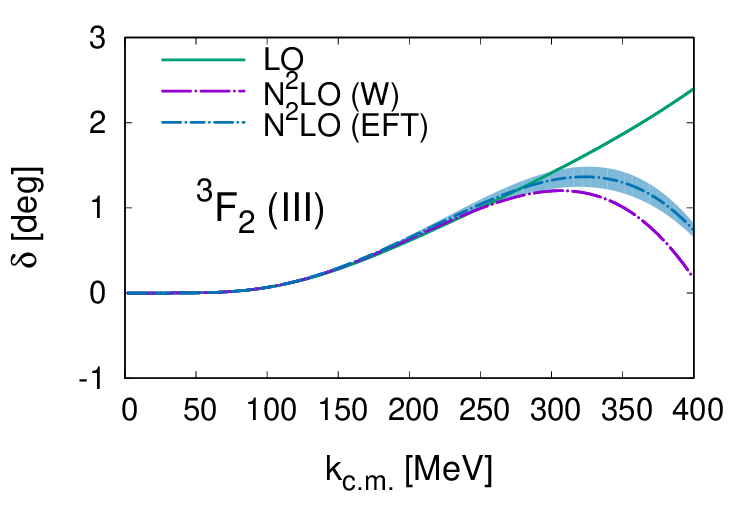}
\end{center}
\caption{Same as Fig.~\ref{fig:W23} but for the ${}^3P_2$, ${}E_2$ and
  ${}^3F_2$ partial waves (nuclear bar phase shifts).
}
\label{fig:W23-3C2}
\end{figure*}


\section{Analyzing EFT potentials that follow the Weinberg counting}

The breakdown of power counting is not universal to all amplitudes
in the Weinberg prescription.
Whether this happens or not actually depends on the interplay between
the regulator and the cutoff, and while in a few cases
there will be severe power counting breakdowns,
in others power counting will work perfectly fine.
Besides, from the initial Weinberg potential of Ray, Ordo\~nez and van Kolck~\cite{Ordonez:1993tn,Ordonez:1995rz}
(which can be understood as a proof of concept of the applicability
of EFT ideas to nuclear physics) till the present day, the EFT descriptions of
the nuclear forces have improved dramatically.
For discussion purposes I will refer to the work of Ray, Ordo\~nez and
van Kolck~\cite{Ordonez:1993tn,Ordonez:1995rz} as the {\it first generation}
of EFT potentials, which contains more or less the same ingredients
as the calculation of Fig.~\ref{fig:W1} (except for the use of
time-ordered perturbation theory and the explicit inclusion of
isobar $\Delta$ excitations in Ref.~\cite{Ordonez:1995rz}).
For this reason, though I have not explicitly analyzed it, I do not expect
the first generation potential of Ref.~\cite{Ordonez:1995rz} to do much
better in terms of power counting than the calculation of Fig.~\ref{fig:W1}.
The first ${\rm N^2LO}$ potential that signaled the quantitative viability of
Weinberg's prescription twenty years ago is the one from Epelbaum, Gl\"ockle
and Mei{\ss}ner~\cite{Epelbaum:2003gr,Epelbaum:2003xx}, which I will refer
to as a {\it second generation} EFT potential and features an energy
independent potential and an improved regularization and
renormalization process owing to the inclusion of
spectral function regularization (SFR).
The more recent ${\rm N^2LO}$ potential of Gezerlis
et al.~\cite{Gezerlis:2014zia} is local and pushes
the SFR cutoff to harder values (which is useful to avoid the long-range
distortions of the chiral potential induced by SFR~\cite{Valderrama:2008kj}).
I will refer to it as a {\it third generation} potential.

Here I analyze the power counting of the aforementioned second and
third generation potentials for the S- and P-wave partial waves
at $\rm N^2LO$.
Two of these partial waves --- $^1S_0$ and $^3P_0$ --- are particularly
interesting because their power counting is expected to deviate
from Weinberg~\footnote{Actually, the ${}^3P_2$-${}^3F_2$
  partial wave is an edge case: at ${\rm LO}$ it is a triplet
  for which the tensor force is attractive and thus it will
  eventually display a strong cutoff dependence that requires
  the modification of the Weinberg counting~\cite{Nogga:2005hy}.
  This problem however is only evident at cutoffs that are considerably
  harder than in the ${}^3P_0$ case (the other attractive P-wave
  triplet), which is the probable reason why I do not observe
  deviations in the power counting of the ${}^3P_2$-${}^3F_2$
  channel with respect to Weinberg.
  Moreover, as argued in~\cite{Wu:2018lai}, the ${}^3P_2$-${}^3F_2$ tensor force
  can be treated perturbatively, which is consistent with the previous
  observation and the results I have obtained.
},
at ${\rm LO}$ for the $^3P_0$ case~\cite{Nogga:2005hy} and at subleading orders
for the $^1S_0$ case~\cite{Valderrama:2009ei,Long:2012ve}.
Thus the naive expectation is that ${}^1S_0$ and ${}^3P_0$ will show signs of
power counting breakdown, while the remaining partial waves ---
${}^3S_1$-${}^3D_1$, ${}^1P_1$, ${}^3P_1$ and
${}^3P_2$-${}^3F_2$ --- will
follow Weinberg's counting.

The result of this analysis for the ${}^1S_0$ and ${}^3P_0$ channels
is shown in Fig.~\ref{fig:W23},
where one can check that power counting is still failing
for the amplitudes of the second generation potential
but this failure is nonetheless less severe
than the one in Fig.~\ref{fig:W1} : now the breakdown scales are
about $M = 310\,{\rm MeV}$ and $M = 270\,{\rm MeV}$ for
the ${}^1S_0$ and ${}^3P_0$, respectively.
There is an improvement of the implicit power counting features of
the third generation potential of Ref.~\cite{Gezerlis:2014zia}
over the second generation one in the particular case of
the $^1S_0$ partial wave, where the breakdown scale
is now about $M = 380\,{\rm MeV}$.
Meanwhile, for the  $^3P_0$ partial wave there is a worsening with
respect to the second generation potential, with the new breakdown
scale being $M = 190\,{\rm MeV}$ (to be compared with
the previous $M = 270\,{\rm MeV}$).
The low breakdown scale in the $^3P_0$ partial wave is mostly
a consequence of mismatches at low momenta. 
In contrast, at moderate momenta, the perturbative reexpansion of
the $^3P_0$ phase shifts works relatively well,
as can be appreciated in Fig.~\ref{fig:W23}.
Yet, the EFT errors of Eqs.~(\ref{eq:err-sum-abridged}) and
(\ref{eq:err-breakdown}) give more weight to the behavior of
the phase shifts at low than at intermediate momenta,
hence explaining the low breakdown scale of the $^3P_0$ channel
despite the good visual appearance of its perturbative reexpansion.

The results for the rest of the partial waves are shown
in Figs.~\ref{fig:W23-3C1} (${}^3S_1$-${}^3D_1$),
~\ref{fig:W23-3P} (${}^1P_1$ and ${}^3P_1$) and
~\ref{fig:W23-3C2} (${}^3P_2$-${}^3F_2$).
I tabulate the perturbative breakdown scales for the S- and P-waves
in Table \ref{tab:W-breakdown}, where I only define $M$
for the channels to which the subleading contacts
are fitted.
That is, the ${}^3D_1$, $E_2$ and ${}^3F_2$ phases are excluded
from Table \ref{tab:W-breakdown} and their truncation errors
use the average of the breakdown scales of the lower angular momentum waves
to which they are coupled (i.e. ${}^3D_1$ uses the average of $M$
from ${}^3S_1$ and $E_1$, while $E_2$ and ${}^3F_2$ use $M$
from the ${}^3P_2$ phase).

It can be appreciated from Table \ref{tab:W-breakdown} that the naive
expectations about the success and failure of the Weinberg counting
are perfectly fulfilled for the second generation potential.
For the third generation one the situation is more open to interpretation,
but still in line with the previous expectations: as previously
mentioned, even though the breakdown scale of the ${}^3P_0$ case
is rather soft (owing to the low momentum behavior of
the phase shifts), its perturbative reexpansion
at moderate momenta seems to work just fine.
The reason might be the local nature of
the Gezerlis et al.~\cite{Gezerlis:2014zia} potential:
while for a non-local potential the ${}^3P_0$ partial wave has a
dedicated contact interaction only acting on it, in a local
potential the isovector triplets share three contacts
(with central, tensor and spin-orbit structures).
Effectively what one sees is that there is a balancing act
on how to fit the low momentum behavior of the ${}^3P_0$,
${}^3P_1$ and ${}^3P_2$ partial waves with
the aforementioned three contacts, a phenomenon which does not happen
in the non-local potential of Refs.~\cite{Epelbaum:2003gr,Epelbaum:2003xx}.
It happens that this balance works against the ${}^3P_0$ channel.
For comparison, had I fitted a central, tensor or spin-orbit counterterm
exclusively to the ${}^3P_0$ phase shifts, the breakdown scale would
have been $M = 233$, $310$ and $380\,{\rm MeV}$, respectively.

The only clear outlier to the expected failure pattern in the Weinberg
counting seems to be the $E_1$ phase shift, which happens to have
a relative soft breakdown scale: $M = 330$ and $120\,{\rm MeV}$
for the second and third generation potentials, respectively.
This might indicate a poor perturbative convergence of observables that
depend on the S- and D-wave mixing, or it might be an artifact of
the criterion by which the breakdown scale is defined:
it should be stressed that Eqs.~(\ref{eq:err-sum-abridged}) and
(\ref{eq:err-breakdown}) are merely a convenient
way to approximate the true breakdown scale (notice
that this is a potential issue not only for $E_1$
but for all partial waves, including $^1S_0$ and $^3P_0$).

The fact that the ${}^1S_0$ breakdown scale (and to a lesser extent
the $^3P_0$ one) is different enough for the second and third
generation potentials might seem somewhat puzzling,
at least at first sight.
After all, every ${\rm N^2LO}$ EFT potential contains the same
physics and should yield the same results modulo truncation errors.
Naively this includes their breakdown scales when the amplitudes are
expanded according to their power counting.
But the Weinberg counting generates cutoff dependent amplitudes, which
is the reason why modifications to it have been proposed.
In terms of the EFT-expanded amplitude this means that its breakdown scale
will be regulator dependent, from which the power counting
properties of individual potentials might in principle
differ a lot.
That is, this is merely a reflection of the non-renormalizability
of the Weinberg counting, only that it manifests at lower
cutoffs (with the specific cutoff at which this happens
being regulator-dependent).

Here it is also interesting to notice the results of the BUQEYE
collaboration~\cite{Furnstahl:2015rha,Melendez:2017phj},
according to which the ${\rm N^3LO}$ and ${\rm N^4LO}$ third generation
potentials~\cite{Epelbaum:2014efa,Epelbaum:2014sza}
seem to follow the Weinberg counting rules within
a certain statistical degree of confidence.
It is sensible to think that there is a relation between
the findings of the BUQEYE collaboration and the fact
that the ${\rm N^2LO}$ third generation potential chosen here
has better power counting features
than the second generation one.
But to fully test this hypothesis one should extend
the potential-vs-amplitude comparison
to higher orders in the Weinberg expansion.
Nonetheless it is interesting to check that Ref.~\cite{Melendez:2017phj}
suggests a breakdown scale of $M = 500\,{\rm MeV}$ and $400\,{\rm MeV}$
for $R_c = 1.1\,{\rm fm}$ and $1.2\,{\rm fm}$,
respectively, though of course this refers to the ${\rm N^3LO}$
and ${\rm N^4LO}$ potentials
of Refs.~\cite{Epelbaum:2014efa,Epelbaum:2014sza}, not necessarily to
the ${\rm N^2LO}$ potential analyzed here~\cite{Gezerlis:2014zia}.
Yet, it is sensible to think that the ${\rm N^2LO}$ breakdown scale
will be similar, which suggests that once the perturbative breakdown
scale of Table \ref{tab:W-breakdown} reaches the
$M = (400-500)\,{\rm MeV}$ range, power counting
can be said to be preserved.

The technical details of the comparison of Fig.~\ref{fig:W23} are similar to
the ones behind the calculation of Fig.~\ref{fig:W1} and can be consulted
in Appendices \ref{app:p-space} and \ref{app:r-space} for
the second and third generation potentials, respectively.
The EFT truncation error is the one described in
Eqs.~(\ref{eq:err-sum-abridged}) and (\ref{eq:err-breakdown}),
which is explained in more detail in Appendix \ref{app:err}.
There are a few details that are worth commenting now.
For the finite-range piece of the EFT potential, i.e. one- and two-pion
exchange, I use the same conventions and regularizations as in the
original potentials~\cite{Epelbaum:2003gr,Epelbaum:2003xx,Gezerlis:2014zia}.
In particular I include SFR for the two-pion exchange piece of
the EFT potential. 
The perturbative amplitude $T_{\rm EFT}$ for the $^1S_0$ partial wave is fitted
to the corresponding ${\rm N^2LO}$ amplitude in the range
$k_{\rm c.m.} = (40-120)\,{\rm MeV}$, where $k_{\rm c.m.}$ is
the center-of-mass momentum.
For the second generation EFT potential, the $^3P_0$ partial wave is fitted
in the $k_{\rm c.m.} = (40-120)\,{\rm MeV}$, while for the rest of the triplets
this range is extended to $k_{\rm c.m.} = (50-150)\,{\rm MeV}$.
For the third generation EFT potential, all the triplets have to be fitted
together within the $k_{\rm c.m.} = (30-90)\,{\rm MeV}$ range.
It is possible to fit for larger momenta, but this does not translate
into better phase shifts when expanding in terms of amplitudes.
On the contrary, in the case of the ${}^1S_0$ and ${}^3P_0$ waves
this only worsens the matching between $T_W$ and $T_{\rm EFT}$.
For the second generation potential
of Refs.~\cite{Epelbaum:2003gr,Epelbaum:2003xx}, which do
not provide a ${\rm LO}$ potential, I fit the contact-range couplings of
the ${}^1S_0$ and ${}^3S_1$ partial waves to the scattering length and
low momentum ($k_{\rm c.m.} = 20\,{\rm MeV}$) Nijmegen II phase shift,
respectively.

\begin{table}
\begin{center}
\begin{tabular}{|c|c|c|}
  \hline \hline
  Partial Wave & $M$ (II) & $M$ (III) \\
  \hline
  ${}^1S_0$ & $310$ & $380$ \\
  ${}^3S_1$ & $530$ & $530$ \\
  $E_1$ & $330$ & $120$ \\
  \hline
  ${}^1P_1$ & $1730$ & $600$ \\
  ${}^3P_0$ & $270$ & $190$ \\
  ${}^3P_1$ & $570$ & $450$ \\
  ${}^3P_2$ & $1670$ & $1250$ \\
  \hline \hline
\end{tabular}
\end{center}
\caption{
  Breakdown scale (in units of ${\rm MeV}$) of the perturbative reexpansion of
  the Weinberg prescription for the second and third generation potentials
  (denoted as II and III in the Table).
  The criterion by which they are calculated is that the amplitude- and
  potential-based expansions are compatible within errors
  for center-of-mass momenta $k_{\rm c.m.} < M$ (with the uncertainties
  defined in Eq.~(\ref{eq:err-sum-abridged})).
  Values are approximated to the closest tens of ${\rm MeV}$.
}
\label{tab:W-breakdown}
\end{table}


\section{Building EFT potentials that preserve power counting}

The perturbative reexpansion of the potentials of Refs.~\cite{Epelbaum:2003gr,Epelbaum:2003xx,Gezerlis:2014zia} fails in the ${}^1S_0$ and ${}^3P_0$ partial waves.
This is interesting in the sense that it provides an independent
confirmation of the existence of the power counting modifications
that have been discussed for two decades in the literature~\cite{Nogga:2005hy,Birse:2005um,Valderrama:2009ei,Valderrama:2011mv,Long:2011qx,Long:2011xw,Long:2012ve}.
The difference though is the methodology used here: instead of relying
on renormalizability (in particular in guaranteeing the cutoff
independence of the amplitudes), the present results make use
of the expectation that perturbative modifications of
the potential are expected to generate perturbative
modifications of the amplitude.

The application of the power counting stemming from RG invariance~\cite{Birse:2005um,Valderrama:2009ei,Valderrama:2011mv,Long:2011qx,Long:2011xw,Long:2012ve} to nucleon-nucleon
scattering is still in the proof of concept phase~\cite{Valderrama:2009ei,Valderrama:2011mv,Long:2011qx,Long:2011xw,Long:2012ve},
a situation analogous to that of Weinberg's counting thirty years ago.
Though originally intended for expanding amplitudes, this power counting
can be used to construct a potential in the same way
as in the case of the Weinberg counting.
Then the non-perturbative amplitudes obtained from this potential
can also be reexpanded following the previous ideas, to check whether
the power counting in the potential and the amplitudes coincide.

There is however a certain degree of ambiguity in the power counting
of pionful EFT, where different works disagree regarding a few specific
details of the ordering of operators~\cite{Birse:2005um,Valderrama:2009ei,Valderrama:2011mv,Long:2011qx,Long:2011xw,Long:2012ve} (a partial explanation
can be found in~\cite{Valderrama:2016koj}).
Given this ambiguity, I have opted to implement a minimal
modification to the Weinberg counting used in Refs.~\cite{Epelbaum:2003gr,Epelbaum:2003xx,Gezerlis:2014zia} in which I only add the counterterms
that are strictly necessary to make the potential- and amplitude-based
expansions compatible (for these potentials and their regularization choices).

I begin with the second generation potential~\cite{Epelbaum:2003gr,Epelbaum:2003xx},
which has the advantage of being non-local: this means that it can be modified
in a partial wave basis.
In particular, at ${\rm LO}$ I only add a contact
in the ${}^3P_0$ partial wave 
\begin{eqnarray}
  \langle p' | \delta V_C^{(\nu = 0)}({}^3P_0) | p \rangle = C_0^{(\nu)}({{}^3P_0})\,p p' \, ,
  \label{eq:delta-Vc-lo-II}
\end{eqnarray}
while at ${\rm NLO}$ and ${\rm N^2LO}$ I include higher-derivative terms
in the ${}^1S_0$ and ${}^3P_0$ contact-range potentials:
\begin{eqnarray}
  \langle p' | \delta V_C^{(\nu = 2,3)}({}^1S_0) | p \rangle &=&
  C_4^{(\nu)}({}^1S_0)\,(p^4 + p'^4) \, ,
  \label{eq:delta-Vc-nlo-IIa}
  \\
  \langle p' | \delta V_C^{(\nu = 2,3)}({}^3P_0) | p \rangle &=&
  C_2^{(\nu)}({{}^3P_0})\,p p'\,(p^2 + p'^2) \, . 
    \label{eq:delta-Vc-nlo-IIb}
\end{eqnarray}
The rest of the partial waves remain unaffected by this modification
of the counting.

\begin{figure*}[ttt]
\begin{center}
  \includegraphics[height=4.5cm]{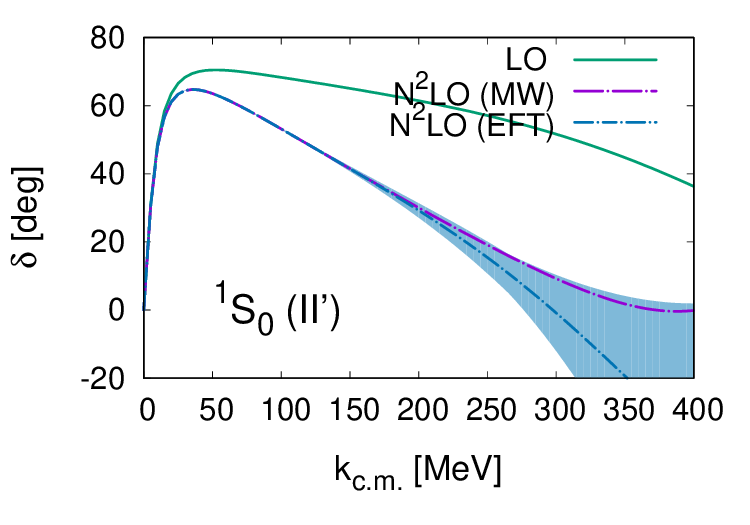}
  \includegraphics[height=4.5cm]{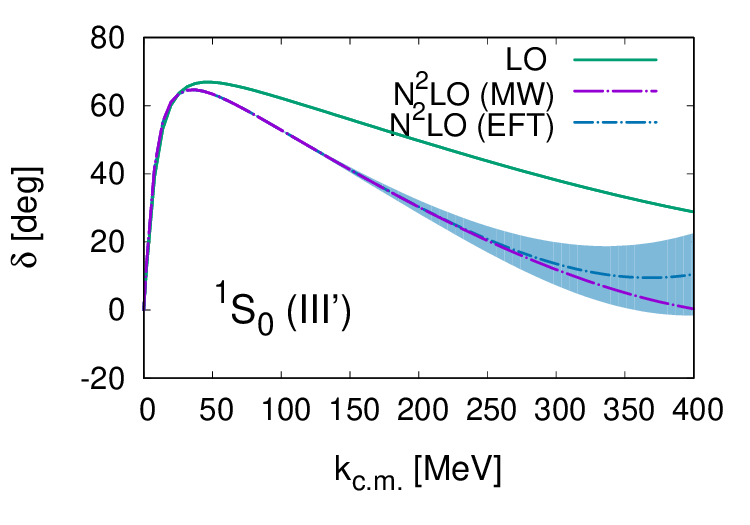}
  \includegraphics[height=4.5cm]{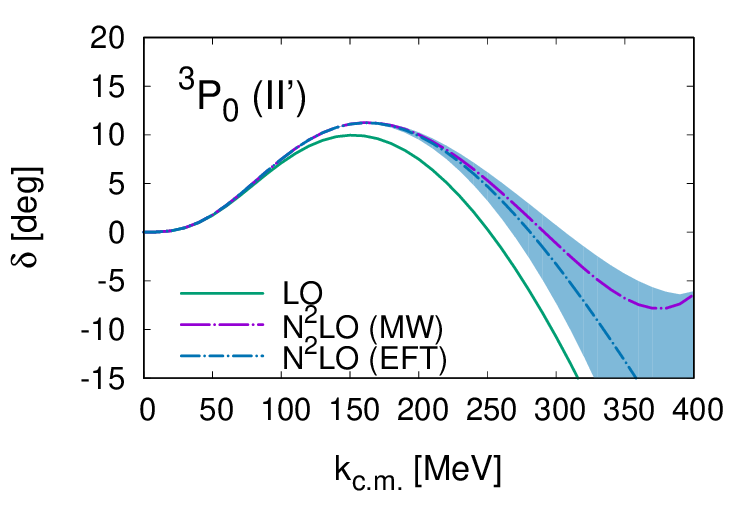}
  \includegraphics[height=4.5cm]{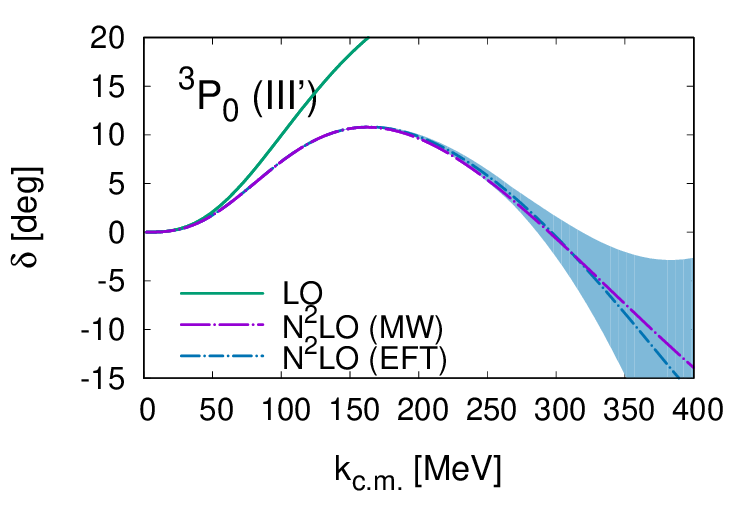}
\end{center}
\caption{Potential (MW, for modified Weinberg, where calculations are
  still fully non-perturbative as in the usual Weinberg prescription)
  and amplitude (EFT, indicating the application of the strict EFT expansion
  for a modified Weinberg counting in the phase shifts) expansions of
  the ${}^1S_0$ and ${}^3P_0$ phase shifts in a modified Weinberg counting
  that improves the agreement between them.
  In this new counting a few counterterms are added to the ${\rm N^2LO}$
  second~\cite{Epelbaum:2003gr,Epelbaum:2003xx} and
  third~\cite{Gezerlis:2014zia} generation
  potentials.
  For the potential of Ref.~\cite{Epelbaum:2003gr,Epelbaum:2003xx},
  a higher order derivative contact-range interactions is included
  in the ${}^3P_0$ case at ${\rm LO}$ (as proposed in Ref.~\cite{Nogga:2005hy})
  while two additional contacts are included at ${\rm NLO}$ in the ${}^1S_0$
  and ${}^3P_0$ waves (in agreement with
  Refs.~\cite{Birse:2005um,Valderrama:2009ei,Long:2011qx,Long:2012ve}).
  For the potential of Ref.~\cite{Gezerlis:2014zia},
  an isovector tensor contact-range interaction is included at ${\rm LO}$,
  which affects the $P$-wave triplets (${}^3P_0$, ${}^3P_1$, ${}^3P_2$),
  while at ${\rm NLO}$ a derivative of the previous tensor contact-range
  interaction is included, plus a four-derivative central counterterm acting
  on the $S=0$, $I=0$ partial waves (i.e. mostly the ${}^1S_0$ channel).
  The roman numerals indicate whether the phases shifts have been calculated
  with the modified second (II') or third (III') generation potential.
  The potential and amplitude expansions are now compatible within
  truncation errors up to momenta of $430$ and $440\,{\rm MeV}$
  ($410$ and $530\,{\rm MeV}$) for the ${}^1S_0$ and ${}^3P_0$
  channels with the second (third) generation potentials,
  respectively.
}
\label{fig:pionful-EFT}
\end{figure*}

For the third generation potential of Ref.~\cite{Gezerlis:2014zia},
which is local, the modifications are more involved:
for a local potential, changes in the contact-range potential
usually affect a few partial waves simultaneously.
In particular there are three possible choices for a triplet
P-wave counterterm: a central, tensor or spin-orbit structure.
They do not only influence the $^3P_0$ partial wave, but also the $^3P_1$
and $^3P_2$ channels (and in the case of the spin-orbit counterterm,
it also affects the ${}^3S_1$-${}^3D_1$ channel, though its
influence is considerably weaker than
for the triplet P-waves).
The choice that provides the best improvement of the power counting
properties of the potential is a tensor counterterm.
Thus, at ${\rm LO}$ the contact-range potential is modified
by including the following term
\begin{eqnarray}
  && \delta V_C^{(\nu = 0)}(\vec{q}\,) =
  \frac{1}{4}\,(3 + \vec{\tau}_1 \cdot \vec{\tau}_2) \nonumber \\
  && \qquad \times
  \,\left( 3\vec{\sigma}_1 \cdot \vec{q}\,\vec{\sigma}_2 \cdot \vec{q}
  - \vec{\sigma}_1 \cdot \vec{\sigma}_2\,q^2\right)\, c_{2t}^{(\nu)}({}^3P)
  \, , \nonumber \\
  \label{eq:delta-Vc-lo-III}
\end{eqnarray}
while at ${\rm NLO}$ and ${\rm N^2LO}$ one simply adds
the higher-derivative version of this counterterm,
plus a second ``central'' contact targeting
the $I=1$, $S=1$ sector (mainly acting on
the $^1S_0$ partial wave):
\begin{eqnarray}
  && \delta V_C^{(\nu = 2,3)}(\vec{q}\,) =
  \frac{1}{4}\,(3 + \vec{\tau}_1 \cdot \vec{\tau}_2) \nonumber \\
  && \qquad \times \Big[
    \frac{1}{4}\,(1 - \vec{\sigma}_1 \cdot \vec{\sigma}_2)\,q^4\,c_{4}^{(\nu)}({}^1S_0) + 
    \nonumber \\ && \qquad
    \,\left( 3\vec{\sigma}_1 \cdot \vec{q}\,\vec{\sigma}_2 \cdot \vec{q}
    - \vec{\sigma}_1 \cdot \vec{\sigma}_2\,q^2\right)\,q^2\, c_{4t}^{(\nu)}({}^3P)    \Big] \, .
  \nonumber \\
  \label{eq:delta-Vc-nlo-III}
\end{eqnarray}
It turns out that the effects of this new isovector tensor
contact are most visible on the ${}^3P_0$ phase shift.

The new potential and amplitude expansions are shown
in Fig.~\ref{fig:pionful-EFT}, where it can be appreciated
a marked improvement of the momenta at which both expansions are equivalent.
For the second generation potential, the new breakdown scales are
$M = 430\,{\rm MeV}$ and $440\,{\rm MeV}$ for the ${}^1S_0$ and
${}^3P_0$, respectively.
Meanwhile for the third generation one, one has $M = 410\,{\rm MeV}$ and
$530\,{\rm MeV}$ instead (i.e. a modest improvement for ${}^1S_0$ and
a large one for ${}^3P_0$).
As previously stated, the third generation potential is local and its new
tensor counterterms are fitted to all the $P$-wave triplets
simultaneously (and affect all of them). Yet it so happens
that the ${}^3P_0$ channel is most influenced by them:
the changes to the $^3P_1$ and $^3P_2$ phase shifts
are relatively minor in comparison.
This is no coincidence though: the ${}^3P_0$ is a triplet for which
the one-pion exchange tensor
force is particularly attractive, which increases its sensitivity to short
range physics. In terms of power counting the lowest order contact-range
interaction becomes ${\rm LO}$ in this partial wave~\cite{Nogga:2005hy},
while the subleading order couplings are enhanced by a factor ranging
from $Q^2$~\cite{Long:2011qx} to $Q^{5/2}$~\cite{Birse:2005um,Valderrama:2011mv}
depending on the assumptions (often implicit) of the analysis made.
In contrast the $^3P_1$ is a repulsive triplet, while for the ${}^3P_2$ channel
the tensor force, though attractive, is considerably weaker than
in the ${}^3P_0$ case (up to the point that one-pion exchange
can be treated as a perturbation in the ${}^3P_2$-${}^3F_2$
channel~\cite{Wu:2018lai}).

\begin{figure}[ttt]
\begin{center}
  \includegraphics[height=4.5cm]{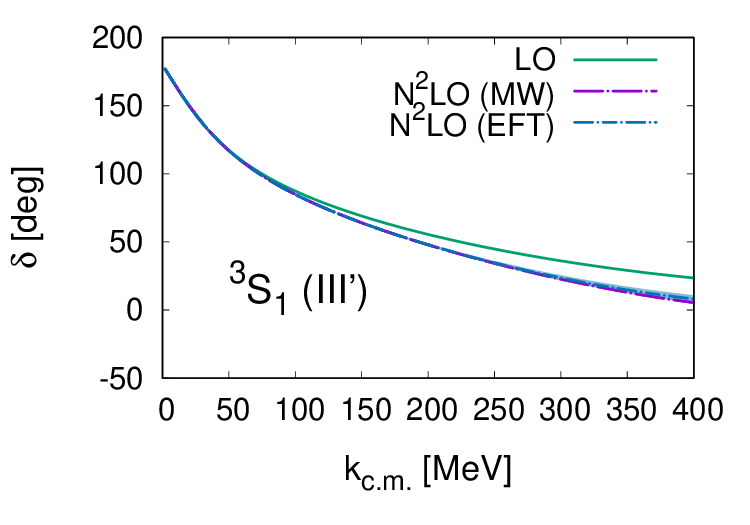}
  \includegraphics[height=4.5cm]{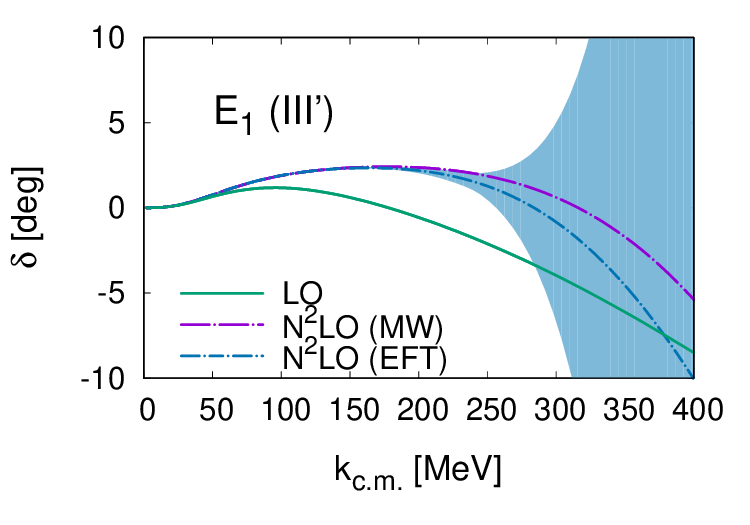}
  \includegraphics[height=4.5cm]{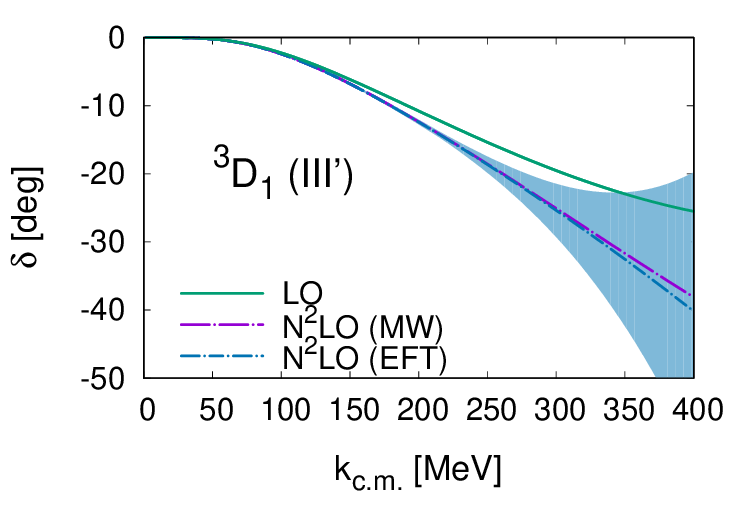}
\end{center}
\caption{Expansions of the ${}^3S_1$-${}^3D_1$ (nuclear bar) phase shifts
  in terms of the potential (MW) and the scattering amplitude (EFT)
  when the power counting is modified.
  Calculations correspond to the third generation
  potential of Ref.~\cite{Gezerlis:2014zia} when additional counterterms
  are included (check Eqs.~(\ref{eq:delta-Vc-lo-III}) and
  (\ref{eq:delta-Vc-nlo-III})) as to improve the power
  counting properties in the $^1S_0$ and $^3P_0$
  channels.
  Even though the ${}^3S_1$-${}^3D_1$ channel is not directly affected by
  the new contacts, it is indirectly affected because of the changes
  in the parameters when refitting the potential.
}
\label{fig:pionful-EFT-3C1}
\end{figure}

\begin{figure}[ttt]
\begin{center}
  \includegraphics[height=4.5cm]{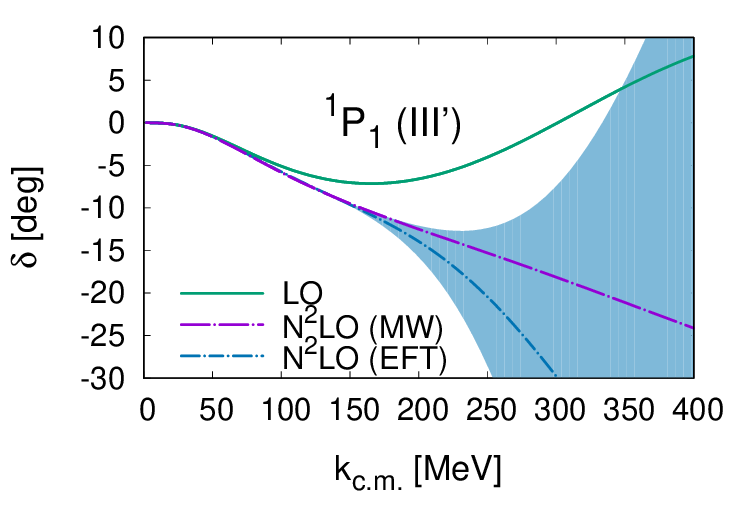}
  \includegraphics[height=4.5cm]{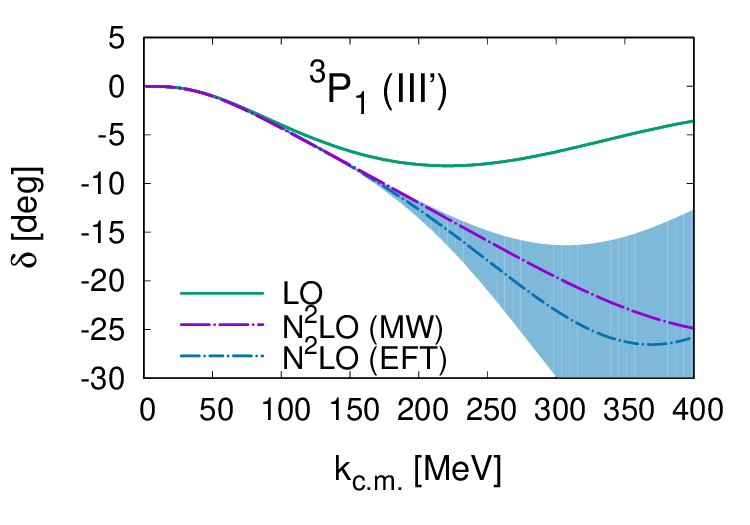}
\end{center}
\caption{Expansions of the ${}^1P_1$ and ${}^3P_1$ phase shifts in terms of
  the potential (MW) and the scattering amplitude (EFT) when the power
  counting is modified.
  They are calculated with the potential of Ref.~\cite{Gezerlis:2014zia}
  when the new isovector tensor counterterms of
  Eqs.~(\ref{eq:delta-Vc-lo-III}) and
  (\ref{eq:delta-Vc-nlo-III}) are added.
  The new terms affect the ${}^1P_1$ partial wave indirectly (by means
  of refitting) and the $^3P_1$ phase directly.
}
\label{fig:pionful-EFT-1P1-3P1}
\end{figure}

\begin{figure}[ttt]
\begin{center}
  \includegraphics[height=4.5cm]{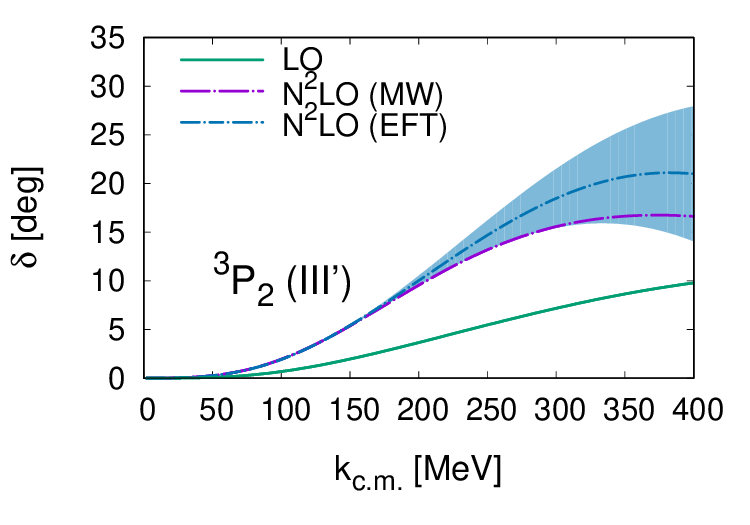}
  \includegraphics[height=4.5cm]{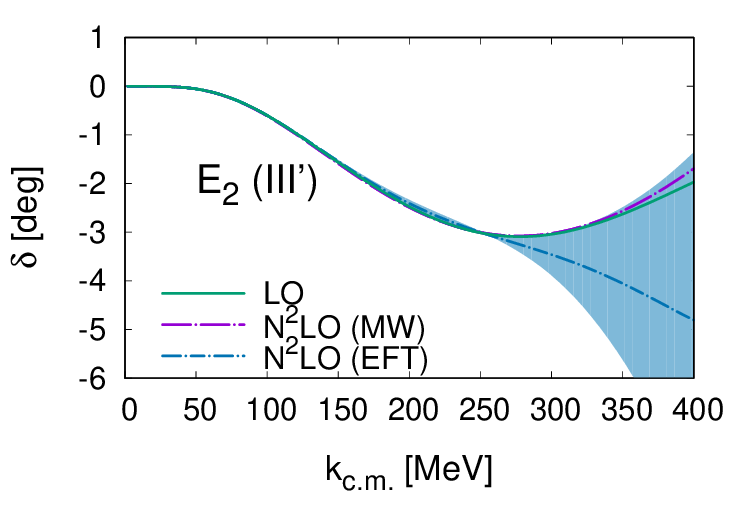}
  \includegraphics[height=4.5cm]{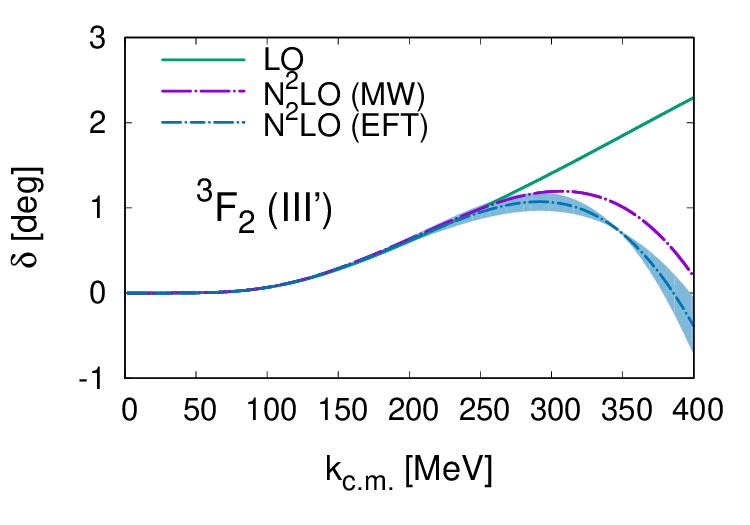}
\end{center}
\caption{Expansions of the ${}^3P_2$-${}^3F_2$ (nuclear bar) phase shifts
  in terms of the potential (MW) and the scattering amplitude (EFT)
  when the power counting is modified.
  The calculations use the potential of Ref.~\cite{Gezerlis:2014zia},
  to which the counterterms of Eqs.~(\ref{eq:delta-Vc-lo-III}) and
  (\ref{eq:delta-Vc-nlo-III}) are added.
}
\label{fig:pionful-EFT-3C2}
\end{figure}

For completeness, I show the other S- and P- phase shifts
for the modified third generation potential
in Figs.~\ref{fig:pionful-EFT-3C1} ($^3S_1$-$^3D_1$),
~\ref{fig:pionful-EFT-1P1-3P1} ($^1P_1$ and $^3P_1$),
and ~\ref{fig:pionful-EFT-3C2} ($^3P_2$-$^3F_2$).
The new breakdown scales of the perturbative reexpansion are listed
in Table \ref{tab:pionful-EFT-breakdown}, where besides the changes
already mentioned for the $^1S_0$ and $^3P_0$ channels
it can be appreciated that for the modified third generation potential
there are small improvements in the $^3P_1$ and $E_1$ waves
and a worsening in $^1P_1$ and $^3P_2$ (though this is not
necessarily significant as their breakdown scales are of
the order of what one would naively expect for nuclear EFT).

\begin{table}
\begin{center}
\begin{tabular}{|c|c|c|}
  \hline \hline
  Partial Wave & $M$ (II') & $M$ (III') \\
  \hline
  ${}^1S_0$ & $430$ & $410$ \\
  ${}^3S_1$ & $530$ & $540$ \\
  $E_1$ & $330$ & $140$ \\
  \hline
  ${}^1P_1$ & $1730$ & $430$ \\
  ${}^3P_0$ & $440$ & $530$ \\
  ${}^3P_1$ & $570$ & $490$ \\
  ${}^3P_2$ & $1670$ & $550$ \\
  \hline \hline
\end{tabular}
\end{center}
\caption{
  Breakdown scale (in units of ${\rm MeV}$) of the perturbative reexpansion of
  the modified power counting defined by the addition of the contact-range
  interactions of Eqs.~(\ref{eq:delta-Vc-lo-II}-\ref{eq:delta-Vc-nlo-III})
  to the Weinberg prescription.
  For the modified non-local potential of
  Refs.~\cite{Epelbaum:2003gr,Epelbaum:2003xx}
  (indicated by II' in the Table) only the breakdown scale of
  the $^1S_0$ and $^3P_0$ partial waves changes with respect
  to Table \ref{tab:W-breakdown}.
  In contrast, the effect of the new additions to the local potential of
  Ref.~\cite{Gezerlis:2014zia} (labeled as III') extends to all partial
  waves, though changes are minor except
  in the $^1S_0$ and $^3P_0$ cases.
}
\label{tab:pionful-EFT-breakdown}
\end{table}

The details of the calculations are analogous to the ones in the previous
section (i.e. the calculations in Figs.~\ref{fig:W23}, \ref{fig:W23-3C1},
\ref{fig:W23-3P} and \ref{fig:W23-3C2}), with the most important difference
being the modification of the ${\rm LO}$, ${\rm NLO}$ and ${\rm N^2LO}$
contact-range potentials as given
in Eqs.~(\ref{eq:delta-Vc-lo-II}-\ref{eq:delta-Vc-nlo-III}).
For the modification of the second generation potential of
Refs.~\cite{Epelbaum:2003gr,Epelbaum:2003xx},
the ${\rm LO}$ contact-interaction in the ${}^3P_0$ channel is determined
by reproducing the Nijmegen II phase shift at $k_{\rm c.m.} = 20\,{\rm MeV}$,
while the subleading contacts in the potential expansion are fitted to
the ${}^1S_0$ and ${}^3P_0$ phase shifts from the Nijmegen PWA up to laboratory
energies of $E_{\rm lab} = 250\,{\rm MeV}$ (the original
potential~\cite{Epelbaum:2003gr,Epelbaum:2003xx}
was fitted up to $100\,{\rm MeV}$,
yet the additional counterterms allow to extend the fit to larger energies
in the ${}^1S_0$ and $^3P_0$ partial waves).
For the modification of the third generation potential of
Ref.~\cite{Gezerlis:2014zia}, for ${\rm N^2LO}$
the $\{ {}^1S_0, {}^1P_1\}$ partial waves are fitted up to $200\,{\rm MeV}$
and the $\{ {}^3S_1, E_1, {}^3P_0, {}^3P_1, {}^3P_2\}$ waves up
to $150\,{\rm MeV}$ (with the original potential~\cite{Gezerlis:2014zia}
fitting singlets and triplets up to $E_{\rm lab} = 150\,{\rm MeV}$).
The new ${\rm LO}$ contact of the third generation potential
is calibrated as to optimize the fit of the amplitude expansion,
i.e. it is treated as a parameter controlling
the EFT convergence.
For the perturbative reexpansion (i.e. amplitude expansion) all the subleading
contacts are fitted in the $k_{\rm c.m.} = (50-150)\,{\rm MeV}$ range,
regardless of which potential or partial wave one is considering.

At this point it is also worth mentioning that there are EFT-inspired
potentials which combine finite-range interactions up to order $Q^3$ with
contact-range interactions up to order $Q^4$, for instance
the well-known chiral potential of Ref.~\cite{Entem:2003ft} or
the more recent potential of Ref.~\cite{Piarulli:2014bda}.
The inclusion of the higher order contact-range operators is reminiscent of
the power counting modifications that have been derived from RG analysis
or perturbative
renormalizability~\cite{Birse:2005um,Valderrama:2009ei,Long:2012ve}.
This type of potential might indeed be an accidental realization of the ideas
promoted in this manuscript about finding a potential that
preserves power counting when iterated.
Yet, for this to be true it would be necessary to define what constitutes
the ${\rm LO}$ of the potential of Refs.~\cite{Entem:2003ft,Piarulli:2014bda}
and to check whether their subleading pieces behave
indeed as perturbations (though this last
point looks very likely).


\section{Conclusions}

I have analyzed whether the scattering amplitudes derived
in the Weinberg prescription do actually follow
the Weinberg counting.
The answer depends on the specific partial wave considered:
for the ${}^1S_0$ and ${}^3P_0$ there is a breakdown of the power counting
when one reexpands the scattering amplitudes perturbatively, except
for low momenta well below the expected range of
applicability of this counting.
In contrast for the rest of the partial waves
the Weinberg counting tends to work well.

Previously the ${}^1S_0$ and ${}^3P_0$ partial waves have been known
not to follow Weinberg's counting on the grounds of
renormalizability~\cite{Nogga:2005hy,Birse:2005um,Valderrama:2009ei,Valderrama:2011mv,Long:2011qx,Long:2011xw,Long:2012ve}.
Here I arrive to the same result with a different method, i.e. by
checking whether the nominally subleading terms in the Weinberg
counting do indeed behave as small (and thus perturbative)
corrections.

It is worth noticing that
particular potentials in the Weinberg prescription can do a
better job than others at preserving their expected power
counting for certain regulator and cutoff choices.
For instance, 
the more recent potentials using the Weinberg
prescription~\cite{Gezerlis:2014zia} fare  better
in certain partial waves
than the older ones~\cite{Epelbaum:2003gr,Epelbaum:2003xx},
up to the point that they might be considered to have
acceptable (though not optimal) power counting properties
in the ${}^1S_0$ singlet (while for the ${}^3P_0$ triplet it could be
argued that this partial wave is at a disadvantage because of
the particular definition I have used
for the EFT uncertainties).
The tool for this analysis are the distorted wave perturbative techniques
that are normally used to construct scattering amplitudes
following an explicit power counting~\cite{Valderrama:2009ei,Valderrama:2011mv,Long:2011qx,Long:2011xw,Long:2012ve}.
Yet, the present analysis is of an exploratory nature and more definite
conclusions will require a thorough analysis including second and higher
order perturbation theory for the two-pion exchange potential,
as well as going beyond ${\rm N^2LO}$ (within the Weinberg counting).

There is a tension between traditional nuclear physics methods
and nuclear EFTs with formal power counting.
The former usually relies on purely non-perturbative methods, while the later
requires that the subleading pieces of the interaction are treated
in perturbation theory and renormalized accordingly.
Here I have explicitly shown that it is possible to construct EFT-inspired
potentials with good power counting properties.
This provides a compromise between EFT and traditional
approaches to nuclear physics.
Of course a necessary condition is to include modifications to the
power counting in line with those that have been already derived
from renormalization~\cite{Nogga:2005hy,Birse:2005um,Valderrama:2009ei,Valderrama:2011mv,Long:2011qx,Long:2011xw,Long:2012ve,Valderrama:2016koj}
(instead of NDA) when building this type of potential.
The price to pay is that exact RG invariance will be broken: the EFT-inspired
potentials I am proposing here most probably are only guaranteed to have
good power counting properties for a particular cutoff range.
Yet, at least for a specific set of cutoffs, these EFT-inspired potentials
can be constructed to be equivalent (within EFT uncertainties) to a
standard EFT expansion at the observable level.
But the advantages derived from RG invariance, in particular the possibility
of a priori error estimations (crucial for precision nuclear physics),
are preserved.

To summarize, EFT-inspired potentials provide a practical compromise
in which a key feature of the EFT approach --- error
estimations --- is preserved, while avoiding distorted
wave perturbation theory methods that are rather
difficult to scale up beyond the two-body sector.

\section*{Acknowledgments}

I would like to thank the organizers of the ECT* workshop
``New ideas in constraining nuclear forces'', for which
the ideas of this manuscript were developed.
I would also like to thank Andreas Ekstr\"om for indirectly inspiring
the idea of power counting preservation as a counterpart of
power counting breakdown.
This work is partly supported by the National Natural Science Foundation
of China under Grant No. 12435007, the Fundamental
Research Funds for the Central Universities and the Thousand
Talents Plan for Young Professionals.

\appendix

\section{Perturbation theory in p-space}
\label{app:p-space}

First I will explain perturbation theory as applied in p-space.
The most prominent application is the EFT analysis of the phase shifts
in the Weinberg counting, in particular the second generation
$\rm N^2LO$ Weinberg potential of
Refs.~\cite{Epelbaum:2003gr,Epelbaum:2003xx}.
For this reason I will use here the power counting convention and
notations of the Weinberg prescription.

\subsection{Expansion of the phase shifts}

The starting point is the Lippmann-Schwinger equation
\begin{eqnarray}
T = V + V\,G_0\,T \, , 
\end{eqnarray}
which I expand according to the power counting of the potential
and the T-matrix, see Eqs.~(\ref{eq:V-exp}) and (\ref{eq:T-exp}).
The resulting leading and subleading order Lippmann-Schwinger equations are
the ones in Eqs.~(\ref{eq:T0}-\ref{eq:T3}).

For uncoupled channels
the relation of the full (on-shell) T-matrix with the phase shifts is given by
\begin{eqnarray}
k \cot{\delta(k)} - i k = - \frac{2 \pi}{\mu}\,\frac{1}{T(k)} \, ,
\end{eqnarray}
which can change depending on how one normalizes the T-matrix.
One can simplify the previous relation to
\begin{eqnarray}
\cot{\delta(k)}  = - \frac{2 \pi}{\mu k}\,{\rm Re}\left[ \frac{1}{T(k)} \right]
= - \frac{2 \pi}{\mu k}\,\frac{1}{K(k)} \, ,
\end{eqnarray}
where $K(k)$ is often called the K-matrix.
Expanding according to the power counting, one obtains at ${\rm LO}$
($\nu = 0$)
\begin{eqnarray}
\cot{\delta^{(0)}(k)} = - \frac{2 \pi}{\mu k}\,\frac{1}{K^{(0)}(k)} \, ,
\end{eqnarray}
while at higher orders ($1 \leq \nu < 4$) one has
\begin{eqnarray}
\frac{\delta^{(\nu)}(k)}{\sin^2{\delta^{(0)}}} =
- \frac{2 \pi}{\mu k}\,\frac{K^{(\nu)}(k)}{{K^{(0)}(k)}^2} \, ,
\end{eqnarray}
where I have used again the K-matrix formalism.

For the coupled channels, I simply take into account that $K(k)$ and
$\cot{\delta}$ are now matrices in the orbital angular
momentum space.
For the $K$-matrix it is convenient to define
\begin{eqnarray}
  {\bf M}(k) =
  \begin{pmatrix}
    M_{aa} & M_{ab} \\
    M_{ba} & M_{bb}
  \end{pmatrix} = -\frac{2 \pi}{\mu k} \frac{1}{{\bf K}(k)} \,
  \, ,
\end{eqnarray}
where the matrix is symmetric ($M_{ab} = M_{ba}$). For the generalization
of $\cot{\delta}$ I use the eigen representation~\cite{PhysRev.86.399}:
\begin{eqnarray}
  {\bf R}(\epsilon)\,
  \begin{pmatrix}
    \cot{\delta_{\alpha}} & 0 \\
    0 &  \cot{\delta_{\beta}}
  \end{pmatrix}\,
  {\bf R}(-\epsilon)
\end{eqnarray}
where $\delta_{\alpha}$ and $\delta_{\beta}$ are the eigen phase shifts and
${\bf R}(\epsilon)$ a rotation matrix given by
\begin{eqnarray}
    {\bf R}(\epsilon) =
    \begin{pmatrix}
    \cos{\epsilon} & -\sin{\epsilon} \\
    \sin{\epsilon} &  \cos{\epsilon}
    \end{pmatrix} \, ,
\end{eqnarray}
with $\epsilon$ the mixing phase.
Expanding the eigen phase shifts according to power counting ($1 \leq \nu < 4$)
one finds:
\begin{eqnarray}
  \frac{\delta_{\alpha}^{(\nu)}(k)}{\sin^2{\delta_{\alpha}^{(0)}}}
  &=& - \Big( M^{(\nu)}_{aa}\,\cos^2{\epsilon^{(0)}}
  + M^{(\nu)}_{bb}\,\sin^2{\epsilon^{(0)}}  \nonumber \\
  && \quad + M^{(\nu)}_{ab}\,\sin{2 \epsilon^{(0)}} \Big) \\
  \frac{\delta_{\beta}^{(\nu)}(k)}{\sin^2{\delta_{\beta}^{(0)}}}
  &=& - \Big( M^{(\nu)}_{aa}\,\sin^2{\epsilon^{(0)}}
  + M^{(\nu)}_{bb}\,\cos^2{\epsilon^{(0)}}  \nonumber \\
  && \quad - M^{(\nu)}_{ab}\,\sin{2 \epsilon^{(0)}} \Big) \\
  \epsilon^{(\nu)} \,(\cot{\delta_{\alpha}^{(0)}} &-& \cot{\delta_{\beta}^{(0)}}) =
  M^{(\nu)}_{ab}\,\cos{2 \epsilon^{(0)}} \nonumber \\
  && \qquad \quad - \frac{1}{2}(M_{aa}^{(\nu)}-M_{bb}^{(\nu)})\,\sin{2 \epsilon^{(0)}}
  \, . \nonumber \\
\end{eqnarray}
For obtaining the expansion of the nuclear bar phase
shifts~\cite{PhysRev.105.302} (which are the ones
I am fitting to and showing in the figures),
I use the well-known conversion formulas:
\begin{eqnarray}
  \delta_{\alpha} + \delta_{\beta} &=& \bar{\delta}_{\alpha} + \bar{\delta}_{\beta}
  \, , \\
  \sin{(\delta_{\alpha} - \delta_{\beta})} &=&
  \frac{\sin{2 \bar{\epsilon}}}{\sin{2 \epsilon}}  \, , \\
  \sin{(\bar{\delta}_{\alpha} - \bar{\delta}_{\beta})} &=&
  \frac{\tan{2 \bar{\epsilon}}}{\tan{2 \epsilon}} \, , 
\end{eqnarray}
and expand them perturbatively.

\subsection{The EFT potential}

The EFT potential contains a finite- and contact-range component
\begin{eqnarray}
  V_{\rm EFT}^{(\nu)} = V_{F}^{(\nu)} + V_{C}^{(\nu)} \, .
\end{eqnarray}
The specific expressions for the $\nu < 4$ finite-range potential
(either with SFR or dimensional regularization (DR)) can be consulted
in Refs.~\cite{Epelbaum:2003gr,Epelbaum:2003xx}.
For the contact-range potential, its (unregularized)
representation reads (after partial wave projection
into the $(js)l$-wave, with $l$, $s$, $j$ the orbital,
spin and total angular momentum):
\begin{eqnarray}
  && \langle p', (js) l' | V_{C} | p, (js) l \rangle \nonumber \\
  && \quad =  p^l {p'}^{l'}
  \sum_{n} C_{2n}((js)ll') \, P_{n}({p'}^2,{p}^2) \nonumber \\
  && \quad = p^l {p'}^{l'}\,
  \Big[ C_0 + C_2 (p^2 + {p'}^2) \nonumber \\
    && \quad + \,C_4 (p^4 + {p'}^4) +
    C_4' p^2 {p'}^2 + \dots \Big] \, ,
  \nonumber \\
\end{eqnarray}
where $p'$ and $p$ are the outgoing and incoming center-of-mass momenta of
the nucleons, $l'$ and $l$ the final and initial angular momenta and
$P_{n}(x,y)$ is a polynomial of order ${n}$ of two variables
such that $P_{n}(x,y) = P_{n}(y,x)$. Notice that in the second
line I have simplified the notation by making the angular
momentum labels implicit.

The couplings $C_{2n}((js)ll')$ are partial wave specific, though instead of
the $(js)ll'$ notation from now on I will specify them with the spectroscopic
notation: $C_0({}^1S_0)$, $C_0({}^3S_1)$, $C_0(E_1)$, etc.
At higher order there are terms that are equivalent in the pionless theory,
for instance $C_4\,(p^4 + p'^4)$ and $C_4'\,p^2 {p'}^2$
as shown in Ref.~\cite{Beane:2000fi}. In the pionful theory
they are suspected to be equivalent, but this has not been proven.
Here I find that calculations work better by including only one of them,
which I choose to be $C_4\,(p^4 + p'^4)$ in the $^1S_0$ channel.
However, this term only appears when one modifies the power counting
in Eq.~(\ref{eq:delta-Vc-nlo-IIa}). In the Weinberg counting
they do not appear though until ${\rm N^3LO}$.

It is also important to notice that the contact-range couplings defined here
differ from the ones in Ref.~\cite{Epelbaum:2003xx} in two
aspects: an overall normalization factor of $4\pi$ and the explicit
addition of contributions from two-pion exchange to the couplings
(which is done for a better comparison with the analogous couplings
deduced from saturation in the boson exchange potentials).
The conversion is straightforward and given (schematically) by
\begin{eqnarray}
  C &=&
  \frac{1}{4 \pi}\,\left[ C' - \delta C_{\rm TPE}' \right] \, , 
\end{eqnarray}
where $C$ is the coupling in the convention used here, $C'$ in the convention
of Ref.~\cite{Epelbaum:2003xx} (i.e. the values listed in Table 4 within
that reference) and $\delta C_{\rm TPE}'$ the corrections coming
from two-pion exchange that can be found in Appendix A of
Ref.~\cite{Epelbaum:2003xx}.
Notice that in Ref.~\cite{Epelbaum:2003xx} couplings with zero and two
derivatives are written as $\tilde{C}({}^{(2s+1)}l_j)$ and ${C}({}^{(2s+1)}l_j)$,
respectively, with ${}^{(2s+1)}l_j$ the partial wave where they act
in the spectroscopic notation.

\subsection{Regularization}

For solving the Lippmann-Schwinger equation or its perturbative expansion,
the EFT potential is regularized as follows
\begin{eqnarray}
  \langle p | V(\Lambda) | p' \rangle &=&
  \langle p | V_{\rm DR / SFR} | p' \rangle\,
  f(\frac{p}{\Lambda})\,f(\frac{p'}{\Lambda}) \, ,
\end{eqnarray}
where I take $f(x) = e^{-x^6}$
(as in Refs.~\cite{Epelbaum:2003gr,Epelbaum:2003xx})
and the original EFT potential,
which might be the dimensional regularization (DR)
or the SFR version, depending on the calculation
(DR is only used in Figs.~\ref{fig:W1} and \ref{fig:extravaganza}).

\subsection{Linear fits}
\label{sub:linear-fits-p}

A very convenient feature of the perturbative corrections to the $T$-matrix
is that they depend linearly on the contact-range couplings.
In particular $T^{(\nu)}$ can be subdivided into a finite- and
contact-range piece
\begin{eqnarray}
  T^{(\nu)} = T_F^{(\nu)} + T_C^{(\nu)} \, ,
\end{eqnarray}
where $T_C^{(\nu)}$ can be written as
\begin{eqnarray}
  T_C^{(\nu)} &=& \sum_{n = 0}^{n_{\rm max}(\nu)} C_{2n}^{(\nu)} T_{C_{2n}}^{(\nu)} \, ,
\end{eqnarray}
with $n_{\rm max}(\nu)$ indicating the largest term included
at a given order.
This property extends to the phase shifts, which can also be written
as the sum of a finite- and contact-range contribution: 
\begin{eqnarray}
  \delta^{(\nu)} = \delta_F^{(\nu)} + \sum_{n = 0}^{n_{\rm max}(\nu)}\,
  C_{2n}^{(\nu)} \delta_{C_{2n}}^{(\nu)} \, .
\end{eqnarray}
This linearity of the subleading contact-range contributions to the phase
shift in turns simplifies the fitting procedure.

It is interesting to notice that for the p-space potential of
Refs.~\cite{Epelbaum:2003gr,Epelbaum:2003xx} the amplitude-based expansion of
the contact-range couplings converges rather well to their values
in the potential-based expansion.
This is a surprising feature: the contact-range couplings are not observable
quantities and thus their perturbative reexpansion is not required to
converge to their non-perturbative values.
For the $^1S_0$ partial wave in the potential-based expansion one has
\begin{eqnarray}
  C_0^{\rm N^2LO}({}^1S_0) &=& -4.623\,{\rm fm}^2 \, , \\
  C_2^{\rm N^2LO}({}^1S_0) &=& +2.687\,{\rm fm}^4 \, , 
\end{eqnarray}
while for the amplitude-based expansion one arrives to
\begin{eqnarray}
  \sum_{\nu=0}^{\nu=3} C_0^{(\nu)}({}^1S_0) &=& -4.649\,{\rm fm}^2 \, , \\
  \sum_{\nu=0}^{\nu=3} C_2^{(\nu)}({}^1S_0) &=& +2.469\,{\rm fm}^4 \, , 
\end{eqnarray}
which are similar.
The full comparison for the counterterms of the Weinberg counting is shown
in Table \ref{tab:couplings-II}, while Table \ref{tab:couplings-II-mod}
displays their values when I modify the counting.
Convergence of the perturbative counterterms into their non-perturbative
values does not appear to be a good indicator of power counting
preservation in the phase shifts: while the $^1S_0$ example above
displays excellent counterterm convergence, it does not show
power counting preservation in the phase shifts.
Conversely, the perturbative convergence of the $C_2$ coupling in the $^3S_1$
partial wave is poor, yet the perturbative reexpansion of the phase shifts
in this channel works relatively well.

\begin{table}[ttt]
  \begin{center}
    \begin{tabular}{|c|c|c|c|c|c|}
      \hline \hline
      Coupling & ${\rm LO}$($\nu=0$) &
      $\nu=2$ & $\nu=3$ &
      $\sum_{\nu=0}^{\nu=3}$ & ${\rm N^2LO}$ \\
      \hline \hline
      $C_0({}^1S_0)$ & $-3.168$ & $-1.289$ & $-0.170$ &
      $-4.627$ & $-4.623$ \\
      $C_2({}^1S_0)$ & $-$ & $+1.347$ & $+1.119$ &
      $+2.461$ & $+2.687$ \\
      \hline \hline
      $C_0({}^3S_1)$ & $-1.333$ & $-3.982$ & $+0.758$ &
      $-4.557$ & $-4.526$ \\
      $10\,C_2({}^3S_1)$ & $-$ & $+3.408$ & $-11.781$ &
      $-8.373$ & $+3.960$ \\
      $10\,C_0(\,\,E_1)$ & $-$ & $-5.446$ & $+4.690$ &
      $-0.756$ & $-4.943$ \\
      \hline \hline
      $10\,C_0({}^1P_1)$ & $-$ & $+2.474$ & $+0.731$ &
      $+3.204$ & $+3.417$ \\
      \hline \hline
      $10\,C_0({}^3P_0)$ & $-$ & $+8.896$ & $-1.923$ &
      $+6.973$ & $+16.150$ \\
      \hline \hline
      $10\,C_0({}^3P_1)$ & $-$ & $-5.891$ & $-2.208$ &
      $-8.100$ & $-7.258$ \\
      \hline \hline
      $10\,C_0({}^3P_2)$ & $-$ & $-7.411$ & $+2.297$ &
      $-5.113$ & $-6.632$ \\
      \hline \hline
    \end{tabular}
  \end{center}
  \caption{Comparison of the contact-range couplings in the amplitude- and
    potential-based expansions for the non-local, momentum space potential
    of Refs.~\cite{Epelbaum:2003gr,Epelbaum:2003xx}:
    the results of the amplitude-based expansion are shown under
    the columns $\nu = 0, 2, 3$, while
    for the potential-based expansion the ${\rm LO}$, ${\rm NLO}$
    and ${\rm N^2LO}$ notation is used (though ${\rm NLO}$ is skipped
    because I am only fitting to the ${\rm N^2LO}$ results).
    The ${\rm LO}$ and $\nu=0$ couplings are the same, as both expansions
    are identical at the lowest order. The $\sum_{\nu=0}^{\nu = 3}$ columns
    show the sum of the first few contributions
    in the amplitude-based expansion. The couplings are in units of
    ${\rm fm}^{2 + 2n + l+l'}$, where $l$ and $l'$ are the initial and final
    orbital angular momentum. When the value of a coupling $C$ is
    small, I show instead $10^a \, C$ with $a$ an exponent.
  }
  \label{tab:couplings-II}
\end{table}

\begin{table}[ttt]
  \begin{center}
    \begin{tabular}{|c|c|c|c|c|c|}
      \hline \hline
      Coupling & ${\rm LO}$($\nu=0$) & $\nu=2$ & $\nu=3$ &
      $\sum_{\nu=0}^{\nu=3}$ & ${\rm N^2LO}$ \\
      \hline \hline
      $C_0({}^1S_0)$ & $-3.168$ & $-1.272$ & $-0.911$ &
      $-5.352$ & $-4.636$ \\
      $C_2({}^1S_0)$ & $-$ & $+1.352$ & $+0.843$ &
      $+2.195$ & $+2.689$ \\
      $10^2\,C_4({}^1S_0)$ & $-$ & $-0.275$ & $+13.079$ &
      $+12.809$ & $-3.407$ \\
      \hline \hline
      $C_0({}^3P_0)$ & $1.322$ & $+0.464$ & $-0.822$ &
      $+0.964$ & $+1.481$ \\
      $10\,C_2({}^3P_0)$ & $-$ & $+0.586$ & $-1.912$ &
      $-1.326$ & $-0.692$ \\
      \hline \hline
    \end{tabular}
  \end{center}
  \caption{Same as Table~\ref{tab:couplings-II} but for the modified
    Weinberg counting defined by Eqs.~(\ref{eq:delta-Vc-lo-II}),
    (\ref{eq:delta-Vc-nlo-IIa}) and (\ref{eq:delta-Vc-nlo-IIb}).
    Only the affected partial waves are shown. The finite-range
    part of the potential is not modified and thus identical
    to that of Refs.~\cite{Epelbaum:2003gr,Epelbaum:2003xx}.
  }
  \label{tab:couplings-II-mod}
\end{table}

\section{Perturbation theory in r-space}
\label{app:r-space}

Here I explain how perturbation theory works in r-space.
The principles are the same as in the p-space case, but the details are more
complex because of the local nature of the contact-range interaction.
I only cover the uncoupled channel case: the extension to coupled
channels can be consulted in Appendix A of Ref.~\cite{Valderrama:2021bql},
which uses similar (though not identical) notations and conventions
as in the present work.
The difference with Appendix A of Ref.~\cite{Valderrama:2021bql} is
that here I follow the (more relaxed) power counting conventions of
the Weinberg prescription, while the previous reference
uses RG-style conventions (e.g. in Ref.~\cite{Valderrama:2021bql}
the ${\rm LO}$ potential is $Q^{-1}$,
while here it is counted as $Q^0$).

\subsection{Perturbative (power counting) series}

The starting point is the reduced Schr\"odinger equation for the $l$-wave
\begin{eqnarray}
  -u_k''(r) + \left[ 2\mu V_{\rm EFT}(r) + \frac{l(l+1)}{r^2} \right]\,u_k(r) =
  k^2 u_k(r) \, , \nonumber \\
\end{eqnarray}
where $u_k$ is the reduced wave function, $V_{\rm EFT}$ the EFT potential,
$k$ the center-of-mass momentum, $l$ the orbital angular momentum and
$\mu$ the reduced mass.
The EFT potential and the wave function are expanded as a series
\begin{eqnarray}
  V_{\rm EFT}(r) &=& \sum_{\nu=0}^{\infty} V^{(\nu)}(r) \, , \\
  u_k(r) &=& \sum_{\nu=0}^{\infty} u^{(\nu)}_k(r) \, , \label{eq:uk-expansion}
\end{eqnarray}
where both expansion begin at $\nu = 0$ (here I am using the conventions
of the Weinberg prescription, in which the ${\rm LO}$ potential
is counted as $Q^0$).
At this point, by making use of a suitable Wronskian identity, I arrive at
the following expression for the cotangent of the phase shift
\begin{eqnarray}
  && \cot{\delta} - \cot{\delta^{(0)}} = \nonumber \\
  && \quad \frac{2\mu}{k}\,\int_0^{\infty} dr\,
  u_k^{(0)}(r)\,(V_{\rm EFT}(r) - V^{(0)}(r))\,u_k(r) \, , \nonumber \\
  \label{eq:cotd-wronskian}
\end{eqnarray}
where both sides of the equation are to be expanded and matched
according to power counting. This will be the formula I will use
for calculating the perturbative corrections to the phase shifts.

By expanding the reduced Schr\"odinger equation I obtain this set of
coupled differential equations:
\begin{eqnarray}
  F^{(0)}\,[ u_k^{(0)}(r) ] &=& 0 \, ,  \\
  F^{(0)}\,[ u_k^{(\nu)}(r) ]
  &=& -2 \mu \sum_{\substack{\nu_1, \nu_2 \geq 0 \\ \nu_1 + \nu_2 = \nu}}\,
  V^{(\nu_1)} u_k^{(\nu_2)} \, , 
\end{eqnarray}
where the differential operator $F^{(0)}$ is simply the ${\rm LO}$ reduced
Schr\"odinger equation
\begin{eqnarray}
  F^{(0)}\,[ u_k ] &=& -u_k^{''} + \left[ 2 \mu V^{(0)} +
    \frac{l(l+1)}{r^2} - k^2 \right]\,u_k \, . \nonumber \\
\end{eqnarray}
I am interested in the EFT series of the phase shifts
\begin{eqnarray}
  \delta(k) = \sum_{\nu = 0} \delta^{(\nu)}(k) \, ,
\end{eqnarray}
which can be calculated iteratively.
One begins with $\nu = 0$ for which the wave function behaves as
\begin{eqnarray}
  u_k^{(0)} \to \cot{\delta^{(0)}}\,\hat{j}_l(kr) - \hat{y}_l(kr) \, ,
\end{eqnarray}
from which one extracts the phase shift $\delta^{(0)}$,
where $\hat{j}_l(x) = x j_l(x)$, $\hat{y}_l(x) = x y_l(x)$
with $j_l(x)$ and $y_l(x)$ the spherical Bessel functions. 
From this, one obtains $\delta^{(1)}$ as
\begin{eqnarray}
  [\cot{\delta}]^{(1)} = \frac{2 \mu}{k}\,I_k^{(1)} \, ,
\end{eqnarray}
which happens to be the first term in
the expansion of Eq.~(\ref{eq:cotd-wronskian}).
Here $[\cot{\delta}]^{(1)}$ can be related to $\delta^{(1)}$ by matching
the expansions 
\begin{eqnarray}
  \cot{\delta} = \cot{( \sum_{\nu = 0} \delta^{(\nu)} )} =
  \sum_{\nu = 0} \, [\cot{\delta}]^{(\nu)} \, ,
  \label{eq:cot-exp}
\end{eqnarray}
leading to
\begin{eqnarray}
  [\cot{\delta}]^{(1)} = - (1 + {\cot^2{\delta^{(0)}}} )\,\delta^{(1)} \, .
\end{eqnarray}
$I_k^{(1)}$ is the {\it perturbative integral}, which for $\nu=1$ is given by
\begin{eqnarray}
  I_k^{(1)} &=&
  \int_0^{\infty} dr\, V^{(1)}\,{u_k^{(0)}}^2 \, .
\end{eqnarray}
From this, one can obtain the asymptotic form ($r \to \infty$) of
the subleading correction to the reduced wave function
at the following order in the EFT expansion
\begin{eqnarray}
  u^{(1)}_k(r) \to [\cot{\delta}]^{(1)}\,\hat{j}_l(k r)  \, ,
\end{eqnarray}
where, by integrating downwards towards $r = 0$, one can compute
the full subleading reduced wave function.
If one assumes that one has $\delta^{(0)}$, $\delta^{(1)}$, $\dots$,
$\delta^{(\nu)}$ and $u_k^{(0)}$, $u_k^{(1)}$, $\dots$, $u_k^{(\nu)}$,
one can obtain the $\nu$-th correction to the phase shift as
\begin{eqnarray}
  [\cot{\delta}]^{(\nu)} = \frac{2\mu}{k}\,I_k^{(\nu)} \, ,
\end{eqnarray}
where $[\cot{\delta}]^{(\nu)}$ is the $\nu$-th term in the expansion
of Eq.~(\ref{eq:cot-exp}). The perturbative integral is defined as
\begin{eqnarray}
  I_k^{(\nu)} &=&
  \int_0^{\infty} dr\,\left[ V^{(\nu)}\,{u_k^{(0)}}^2 +
    V^{(\nu-1)} u_k^{(0)} u_k^{(1)} + \dots \right] \nonumber \\
  &=& \int_0^{\infty} dr\,u_k^{(0)} \sum_{\substack{\nu_1, \nu_2 \geq 0 \\ \nu_1+\nu_2 = \nu}}
  V^{(\nu_1)}\,u_k^{(\nu_2)} \, ,
\end{eqnarray}
from which one obtains the asymptotic form of the $u_k^{(\nu)}$
reduced wave function as
\begin{eqnarray}
  u^{(\nu)}_k \to [\cot{\delta}]^{(\nu)} \,\hat{j}_l(k r) \, ,
\end{eqnarray}
and then continue to the next order.
It is important to notice that here one is EFT-expanding all quantities.
Thus the $\nu \geq 1$ perturbative contribution to the phase shifts are
\begin{eqnarray}
  {\delta}^{(1)} &=& - \sin^2{\delta^{(0)}}\,[\cot{\delta}]^{(1)} \, , \label{eq:d1} \\
  {\delta}^{(2)} &=& - \sin^2{\delta^{(0)}}\,[\cot{\delta}]^{(2)}
  + \cot{\delta^{0}}\,(\delta^{(1)})^2 \, , \\
  {\delta}^{(3)} &=& - \sin^2{\delta^{(0)}}\,[\cot{\delta}]^{(3)}
  -\frac{1}{3} (1+ 3\,{\cot^2{\delta^{(0)}}})\,(\delta^{(1)})^3 \nonumber \\
  &+&  2\,\cot{\delta^{(0)}}\,\delta^{(1)}\,\delta^{(2)} \, , \label{eq:d3} 
\end{eqnarray}
plus increasingly involved expressions at higher orders.
However, in the Weinberg counting there are no
corrections to the potential at order $Q$ or $\nu =1$.
In practice this means that only first order perturbation
theory will be used for $1 \leq \nu < 4$.

As previously mentioned, the extension of the previous formulas to coupled
channels is  explained in detail in Appendix A of
Ref.~\cite{Valderrama:2021bql}, the only difference being the power
counting conventions (which are easy to sort out).

\subsection{Are perturbative phase shifts uniquely defined?}

A common misconception is that the perturbative expansion of the
phase shifts is not well-defined and rely on some sort of unitarization.
In fact at first sight it looks like the definition of the perturbative
phase shifts depends on the asymptotic normalization of the wave function.
But this is merely apparent. If instead of the asymptotic normalization
\begin{eqnarray}
  u_k \to \cot{\delta}\,\hat{j}_l(kr) - \hat{y}_l(kr) \, ,
\end{eqnarray}
one had used other normalization, for example
\begin{eqnarray}
  u_k \to \hat{j}_l(kr) - \tan{\delta}\,\hat{y}_l(kr) \, ,
\end{eqnarray}
the final calculation of $\delta^{(\nu)}$ would have been the same (though
it requires a bit of patience to check this explicitly at higher orders).

Ultimately the uniqueness of the $\delta^{(\nu)}$ can be traced back to
the properties of the coefficients in the Taylor or asymptotic series.
In quantum mechanics a perturbative expansion is basically
an expansion on a parameter $\lambda$
\begin{eqnarray}
  \delta(k) = \sum_{\nu = 0}^{\nu_{\rm max}} \lambda^{\nu}\, \hat{\delta}^{(\nu)}(k)
  + \mathcal{O}(\lambda^{\nu_{\rm max} + 1})
  \, ,    
\end{eqnarray}
where $\lambda$ represents a coupling constant (or a ratio of scales
if one is working with an EFT), and $\hat{\delta}^{(\nu)}$
is simply $\delta^{(\nu)}$ rescaled.
In a few instances this expansion will be a Taylor series and
converge below some critical value of the expansion parameter,
i.e. $|\lambda| < \lambda_{\rm crit}$, while in other cases
it will be an asymptotic expansion.
Be it as it may, the coefficients of these two types of expansions
are unique for a specific choice of the expansion
parameter~\cite{Apostol,Erdelyi}, which implies
the uniqueness of $\delta^{(\nu)}$.

The confusion about the non-uniqueness of the perturbative phase shifts
probably refers instead to the exact unitarization of these phase shifts.
Indeed the basic block around which one is expanding is
not the phase shift but its cotangent, see Eq.~(\ref{eq:cot-exp}).
What I am doing in this manuscript is to match
the power counting expansion of these two quantities
\begin{eqnarray}
  \sum_{\nu=0}^{\nu_{\rm max}} \delta^{(\nu)} \leftrightarrow
      {\rm atan}
      {\left( \frac{1}{\sum_{\nu=0}^{\nu_{\rm max}} [\cot{\delta}]^{(\nu)}}\right)}
      \, ,
\end{eqnarray}
where the matching is done {\it order-by-order}.
For $\nu \leq 3$ I have written down the explicit result of this procedure
in Eqs.~(\ref{eq:d1}-\ref{eq:d3}).
In principle it is possible to {\it exactly unitarize} the phase shifts,
e.g. Ref.~\cite{Yao:2016vbz} proposes two possible schemes.
But this is not what I am doing here, where I am content
with perturbative unitarity. Thus all quantities are
being expanded consistently.

\subsection{Regularization of the finite-range potential}

The finite-range potential is regularized as in Ref.~\cite{Gezerlis:2014zia},
that is, one multiplies the EFT potential by
\begin{eqnarray}
  V_{\rm EFT}(r; R_c) = (1 - f(r; R_c))\,V_{\rm EFT}(r) \, ,
\end{eqnarray}
with $f = e^{-(r/R_c)^{2n}}$.
For the power counting analysis of the potential of
Ref.~\cite{Gezerlis:2014zia} one uses $n=2$ and
in addition the EFT potential is further regularized with SFR.
The $n=2$ exponent is not arbitrary:
even though the $Q^2$ and $Q^3$ EFT potentials diverge as $1/r^5$ and $1/r^6$
as $r \to 0$ in dimensional regularization, in SFR they do as $1/r^3$
(see Ref.~\cite{Valderrama:2008kj} for a detailed comparison
between dimensionally regularized and SFR potentials).
This means that $n=2$ is the smallest exponent
that leads to a regularized potential
that is finite at the origin.
For dimensional regularization the required exponent would be $n=3$.

\subsection{The contact-range potential}

The potential of Ref.~\cite{Gezerlis:2014zia} is local (except for
the contact-range spin-orbit interaction, which though local
in r-space is non-local in p-space).
This means that the representation of its contact-range potential
is different.
In momentum space the ${\rm LO}$ contact-range potential is given by
\begin{eqnarray}
  V_C^{(\nu = 0)}(\vec{q}\,) &=& \frac{1}{4}\,\left( 1 - \vec{\sigma}_1 \cdot \vec{\sigma}_2 \right)\,c_0^{(\nu)} ({}^1S_0) \nonumber \\
  &+& \frac{1}{4}\,\left( 3 + \vec{\sigma}_1 \cdot \vec{\sigma}_2 \right)\,
  c_0^{(\nu)} ({}^3S_1) \, , \label{eq:Vc-lo-local}
\end{eqnarray}
which is the sum of a singlet and triplet components.
The couplings indicate the lowest orbital momentum partial wave they affect ---
the $^1S_0$ and $^3S_1$ for the singlet and triplet components, respectively ---
though their effect extends to all the singlets and triplets.
That is, the $c_0({}^1S_0)$ coupling acts not only in $^1S_0$ but also
in $^1P_1$, $^1D_2$ and so on. Analogously $c_0({}^3S_1)$ acts on
$^3S_1$, $^3P_{0,1,2}$, $^3D_{1,2,3}$, etc.

For the ${\rm NLO}$ and ${\rm N^2LO}$ corrections I express them again
as a sum of a singlet and triplet contributions:
\begin{eqnarray}
  V_C^{(\nu = 2,3)}(\vec{q}\, , \vec{k}\,) &=& \frac{1}{4}\,
  \left( 1 - \vec{\sigma}_1 \cdot \vec{\sigma}_2 \right)\,V_{C,s}^{(\nu)}(\vec{q}\,)
  \nonumber \\
  &+& \frac{1}{4}\,\left( 3 + \vec{\sigma}_1 \cdot \vec{\sigma}_2 \right)\,
  V_{C,t}^{(\nu)}(\vec{q}\,, \vec{k}\,) \, , \nonumber \\
\end{eqnarray}
where the singlet contribution is given by
\begin{eqnarray}
  && V_{C,s}^{(\nu = 2,3)}(\vec{q}\,) =
    c_0^{(\nu)}({}^1S_0) \nonumber \\
    && \quad + \frac{1}{4}\,\left( 3 + \vec{\tau}_1 \cdot \vec{\tau}_2 \right)
    {\vec{q}\,}^2\,c_2^{(\nu)} ({}^1S_0) \nonumber \\
    && \quad + \frac{1}{4}\,\left( 1 - \vec{\tau}_1 \cdot \vec{\tau}_2 \right)
    {\vec{q}\,}^2\,c_2^{(\nu)} ({}^1P_1)  \, ,
\end{eqnarray}
and the triplet by
\begin{eqnarray}
  && V_{C,t}^{(\nu = 2,3)}(\vec{q}\,, \vec{k}\,) =
  c_0^{(\nu)}\, ({}^3S_1) \nonumber \\
    && \quad + \frac{1}{4}\,\left( 1 - \vec{\tau}_1 \cdot \vec{\tau}_2 \right)\,
    {\vec{q}\,}^2\,c_2^{(\nu)}\, ({}^3S_1) \nonumber \\
    && \quad + \frac{1}{4}\,\left( 3 + \vec{\tau}_1 \cdot \vec{\tau}_2 \right)\,
    \Big[ \, {\vec{q}\,}^2\,c_2^{(\nu)} ({}^3P)  \nonumber \\
      && \quad + S_{12}(\hat{q})\,{\vec{q}\,}^2\,
      c_{2t}^{(\nu)}({}^3P) \Big] \nonumber \\
    && \quad + i\,(\vec{\sigma}_1 + \vec{\sigma}_2) \cdot (\vec{q}\, \times \vec{k})\, c_{2ls}^{(\nu)}({}^3P) \, ,
    \label{eq:Vc-nlo-local-b}
\end{eqnarray}
where $S_{12}(\hat{q})$ is the tensor operator
\begin{eqnarray}
  S_{12}(\hat{q}) = 3\,\vec{\sigma}_1 \cdot \hat{q}\,\vec{\sigma}_2 \cdot \hat{q}
  - \vec{\sigma}_1 \cdot \vec{\sigma}_2 \, .
\end{eqnarray}
Again, the couplings indicate the lowest order partial wave they affect,
with $^3P = \{ {}^3P_0, {}^3P_1, {}^3P_2 \}$. The spin-orbit contact
also acts on the ${}^3S_1$-${}^3D_1$ coupled channel, though
indirectly by means of the D-wave.

The r-space version of this contact-range potential is given by
\begin{eqnarray}
  V_C^{(\nu = 0)}(\vec{r}\,) &=& \frac{1}{4}\,\left( 1 - \vec{\sigma}_1 \cdot \vec{\sigma}_2 \right)\,c_0^{(0)} ({}^1S_0)\,\delta_{c0}(\vec{r}\,) \nonumber \\
  &+& \frac{1}{4}\,\left( 3 + \vec{\sigma}_1 \cdot \vec{\sigma}_2 \right)\,
  c_0^{(0)} ({}^3S_1)\,\delta_{c0}(\vec{r}\,) \, , \nonumber \\
\end{eqnarray}
at ${\rm LO}$, while at subleading orders one has
\begin{eqnarray}
  && V_{C,s}^{(\nu = 2,3)}(\vec{r}\,) =
    c_0^{(\nu)}({}^1S_0)\,\delta_{c0}(\vec{r}\,) \nonumber \\
    && \quad + \frac{1}{4}\,\left( 3 + \vec{\tau}_1 \cdot \vec{\tau}_2 \right)
    \,c_2^{(\nu)}({}^1S_0)\,\delta_{c2}(\vec{r}\,)  \nonumber \\
    && \quad + \frac{1}{4}\,\left( 1 - \vec{\tau}_1 \cdot \vec{\tau}_2 \right)
    \,c_2^{(\nu)} ({}^1P_1)\,\delta_{c2}(\vec{r}\,)  \, ,
\end{eqnarray}
for the singlet component and
\begin{eqnarray}
  && V_{C,t}^{(\nu = 2,3)}(\vec{r}\,) = c_0^{(\nu)}\, ({}^3S_1)\,\delta_{c0}(\vec{r}\,) \nonumber \\
    && \quad + \frac{1}{4}\,\left( 1 - \vec{\tau}_1 \cdot \vec{\tau}_2 \right)\,
    \,c_2^{(\nu)}\, ({}^3S_1)\,\delta_{c2}(\vec{r}\,) \nonumber \\
    && \quad + \frac{1}{4}\,\left( 3 + \vec{\tau}_1 \cdot \vec{\tau}_2 \right)\,
    \Big[ \,c_2^{(\nu)} ({}^3P)\,\delta_{c2}(\vec{r}\,)  \nonumber \\
      && \quad + S_{12}(\hat{r})\,
      c_{t2}^{(\nu)}({}^3P)\,\delta_{t2}(\vec{r}\,) \Big] \nonumber \\
    && \quad + \vec{L} \cdot \vec{S}\, c_{ls2}^{(\nu)}({}^3P)
    \,\delta_{ls2}(\vec{r}\,) \, ,
\end{eqnarray}
for the triplet component, where $\vec{L}$ and $\vec{S}$ are the total orbital
and spin angular momentum operators and $S_{12}(\hat{r})$ the tensor
operator in r-space.

The Dirac-delta terms are given by
\begin{eqnarray}
  \delta_{c0}(\vec{r}\,) &=& \delta(\vec{r}\,) \, , \\
  \delta_{c2}(\vec{r}\,) &=& (-\Delta)\,\delta(\vec{r}\,) \nonumber \\
  &=& - \left( \frac{d^2}{dr^2} + \frac{2}{r}\,\frac{d}{dr}\right)\,
  \delta(\vec{r}\,) \\
  \delta_{t2}(\vec{r}\,) &=&
  -\left( \frac{d^2}{dr^2} - \frac{1}{r}\,\frac{d}{dr} \right)\,
  \delta(\vec{r}\,) \, , \\
  \delta_{ls2}(\vec{r}\,) &=& \frac{1}{r}\,\frac{d}{dr}\,\delta(\vec{r}\,) \, ,
\end{eqnarray}
with $\Delta$ the Laplacian operator. For their regularization
one makes the substitution
\begin{eqnarray}
  \delta_{c0}(\vec{r}\,) \rightarrow \delta_{c0}(r; R_c) \, ,
\end{eqnarray}
with the regularized Dirac-delta given by
\begin{eqnarray}
  \delta_{c0}({r}; R_c) = \frac{e^{-(r/R_c)^{2n}}}
        {\frac{4}{3} \pi\,\Gamma(1+\frac{3}{2n})\,R_c^3} \, ,
        \label{eq:regul-gaussian}
\end{eqnarray}
that is, a local Gaussian regulator.
Here $R_c$ is the cutoff radius and $n$ coincides with the exponent used
for regularizing the finite-range potential (i.e. $n=2$ for the EFT potential
of Ref.~\cite{Gezerlis:2014zia}).

It is worth noticing that here the representation of the contact-range potential
is spectroscopic-based, where the choice of couplings and operators is different
(though equivalent) to the one originally used in Ref.~\cite{Gezerlis:2014zia}.
Yet, it is the same contact-range potential and the relations between the
couplings in both representations can be worked out easily, yielding
\begin{eqnarray}
  c_0({}^1S_0) &=& d_{11} \, , \\
  c_0({}^3S_1) &=& d_{22}\, , \\
  c_2({}^1S_0) &=& d_3 - d_7 \, , \\
  c_2({}^1P_1) &=& d_5 - d_2 \, , \\
  c_2({}^3S_1) &=& d_1 + \frac{d_2}{3} \, , \\
  c_{t2}({}^3S_1) &=& \frac{d_2}{3} \, , \\
  c_2({}^3P) &=& d_4 + \frac{d_7}{3} \, , \\
  c_{t2}({}^3P) &=& \frac{d_7}{3} \, , \\
  c_{ls2}({}^3P) &=& d_6 \, ,
\end{eqnarray}
where $d_{11}$, $d_{22}$ and $d_{1, \dots, 7}$ are the spectroscopic
low energy constants (i.e. couplings) defined in
Appendix A of Ref.~\cite{Gezerlis:2014zia}.

Finally, for the modified subleading contact-range potential of
Eq.~(\ref{eq:delta-Vc-nlo-III}), its Fourier transform reads:
\begin{eqnarray}
      && \delta V_C^{(\nu = 2,3)}(\vec{r}\,) =
    \frac{1}{4}\,(3 + \vec{\tau}_1 \cdot \vec{\tau}_2) \nonumber \\
    && \qquad \times \Big[
      \frac{1}{4}\,(1 - \vec{\sigma}_1 \cdot \vec{\sigma}_2)\,\,c_{4}^{(\nu)}({}^1S_0)\,\delta_{c4}(\vec{r}\,) + 
      \nonumber \\ && \qquad \quad
      \,S_{12}(\hat{r})\, c_{4t}^{(\nu)}({}^3P)\,\delta_{t4}(\vec{r}\,)
      \Big] \, ,
    \nonumber \\
\end{eqnarray}
where
\begin{eqnarray}
  \delta_{c4}(\vec{r}\,) &=& {(-\Delta)}^2\,\delta(\vec{r}\,) \nonumber \\
  &=& \left( \frac{d^4}{dr^3} + \frac{4}{r}\,\frac{d^4}{dr^3} \right)\,
  \delta(\vec{r}\,) \, , \\
  \delta_{t4}(\vec{r}\,) &=& -\left( \frac{d^2}{dr^2} - \frac{1}{r}\,\frac{d}{dr} \right)\,\,\delta_{c2}(\vec{r}\,) \nonumber \\
  &=& +\left( \frac{d^4}{dr^3} + \frac{1}{r}\,\frac{d^3}{dr^3}
  - \frac{6}{r^2}\,\frac{d^2}{dr^2} + \frac{6}{r^3}\,\frac{d}{dr} \right)\,
  \delta(\vec{r}\,) \, . \nonumber \\
\end{eqnarray}
For its regularization, one simply substitutes $\delta(\vec{r}\,)$ by
$\delta_{c0}(r; R_c)$ again.

\subsection{Linear fits}
\label{subsec:linear-fits}

The previous description of the calculation of the perturbative phase shifts
assumes that the EFT potential is perfectly well-known.
As has been already discussed, the contributions to the EFT potential
can be further subdivided into a finite- and contact-range piece
\begin{eqnarray}
  V^{(\nu)} = V_F^{(\nu)} + V_C^{(\nu)} \, ,
\end{eqnarray}
where the finite-range piece is usually well-known,
but the contact-range piece contains a series of couplings
that has to be determined from available data
\begin{eqnarray}
  V_c^{(\nu)} = \sum_{n = 0}^{n_{\rm max}(\nu)} C_{2n}^{(\nu)}\,V_{C_{2n}}^{(\nu)} \, ,
\end{eqnarray}
where $n_{\rm max}(\nu)$ indicates where to cut the derivative
expansion at each order and $C_{2n}^{(\nu)}$ is meant as
an abstract representation of the contact-range
couplings, not its concrete representation
that can be found in Eqs.~(\ref{eq:Vc-lo-local}-\ref{eq:Vc-nlo-local-b}).
This property translates directly into the structure of
the perturbative integrals
\begin{eqnarray}
  I_k^{(\nu)} = I_F^{(\nu)} +
  \sum_{n = 0}^{n_{\rm max}(\nu)} C_{2n}^{(\nu)} I_{C_{2n}}^{(\nu)} \, ,
\end{eqnarray}
and the phase shifts too
\begin{eqnarray}
  \delta^{(\nu)} = \delta_F^{(\nu)} +
  \sum_{n = 0}^{n_{\rm max}(\nu)} C_{2n}^{(\nu)} \delta_{C, 2n}^{(\nu)} \, .
\end{eqnarray}
This implies that the determination of the couplings $C_{2n}^{(\nu)}$ is easier
than in the non-perturbative case because the observable quantities depend
linearly on these couplings.
The fitting procedure is therefore relatively simple.

In contrast to what happens in the p-space potential of Ref.~\cite{Epelbaum:2003gr,Epelbaum:2003xx} (see Appendix \ref{sub:linear-fits-p} and the discussion
there), for the r-space potential of Ref.~\cite{Gezerlis:2014zia}
the amplitude-based expansion of the contact-range couplings
in general does not converge well to their
non-perturbative values (though there are exceptions).
For the ${}^1S_0$ partial wave in the Weinberg counting one finds
the non-perturbative values
\begin{eqnarray}
  c_0^{{\rm N^2LO}}({}^1S_0) &=&  +3.155\,{\rm fm}^2\, , \\
  10\,c_2^{{\rm N^2LO}}({}^1S_0) &=&  +1.780\,{\rm fm}^4\, ,
\end{eqnarray}
while the perturbative reexpansion yields
\begin{eqnarray}
  \sum_{\nu=0}^{\nu=3}\,c_0^{(\nu)}({}^1S_0) &=&  +2.915\,{\rm fm}^2\, , \\
  10\,\sum_{\nu=0}^{\nu=3}\,c_2^{(\nu)}({}^1S_0) &=& -0.037\,{\rm fm}^4\, ,
\end{eqnarray}
which converges rather poorly for $c_2({}^1S_0)$ (though acceptably
well for $c_0({}^1S_0)$).
Results for all the counterterms are shown
in Table \ref{tab:couplings-III}.
Despite this, the power counting preservation properties of
the r-space potential of Ref.~\cite{Gezerlis:2014zia} are somewhat better
than for the p-space Weinberg potentials~\cite{Epelbaum:2003gr,Epelbaum:2003xx}.
This preservation of the counting is improved further in the modified
power counting, for which the counterterms are listed
in Table \ref{tab:couplings-III-mod},
though the convergence of the couplings still remains case dependent.
Yet, counterterms are not observable and their perturbative convergence
is simply a nice feature (when it happens): what matters is the expansion
of observable quantities.

\begin{table}[ttt]
  \begin{center}
    \begin{tabular}{|c|c|c|c|c|c|}
      \hline \hline
      Coupling & ${\rm LO}$($\nu=0$) & $\nu=2$ & $\nu=3$ &
      $\sum_{\nu=0}^{\nu=3}$ & ${\rm N^2LO}$ \\
      \hline \hline
      $c_0({}^1S_0)$ & $-2.066$ & $-0.049$ & $+5.030$ &
      $+2.915$ & $+3.155$ \\
      $10\,c_2({}^1S_0)$ & $-$ & $+4.848$ & $-4.885$ &
      $-0.037$ & $+1.780$ \\
      $10\,c_2({}^1P_1)$ & $-$ & $-5.337$ & $+9.472$ &
      $+4.135$ & $+0.673$ \\
      \hline \hline
      $c_0({}^3S_1)$ & $-1.040$ & $-0.587$ & $+4.644$ &
      $+3.108$ & $+4.131$ \\
      $10\,c_2({}^3S_1)$ & $-$ & $+2.262$ & $-5.490$ &
      $-3.228$ & $-3.890$ \\
      $10\,c_{t2}({}^3S_1)$ & $-$ & $+1.722$ & $+1.151$ &
      $+2.873$ & $+3.621$ \\
      \hline \hline
      $10\,c_2({}^3P)$ & $-$ & $+1.173$ & $-2.629$ &
      $-1.455$ & $-1.061$ \\
      $10^2\,c_{t2}({}^3P)$ & $-$ & $-2.054$ & $-5.867$ &
      $-7.921$ & $-3.929$ \\
      $10\,c_{ls2}({}^3P)$ & $-$ & $+6.611$ & $-0.119$ &
      $+6.492$ & $+10.041$ \\
      \hline \hline
    \end{tabular}
  \end{center}
  \caption{Same as Table \ref{tab:couplings-II}, but for the 
    the local, coordinate space potential
    of Ref.~\cite{Gezerlis:2014zia}.
    The couplings are in units of ${\rm fm}^{2 + 2n}$, where $n$ refers to
    the number of derivatives of the operator involved (notice that
    the value of $n$ is already shown in the subindex of
    the coupling, e.g. $c_{ls2}$ has $n=2$).
  }
  \label{tab:couplings-III}
\end{table}

\begin{table}[ttt]
  \begin{center}
    \begin{tabular}{|c|c|c|c|c|c|}
      \hline \hline
      Coupling & ${\rm LO}$($\nu=0$) & $\nu=2$ & $\nu=3$ &
      $\sum_{\nu=0}^{\nu=3}$ & ${\rm N^2LO}$ \\
      \hline \hline
      $c_0({}^1S_0)$ & $-2.066$ & $-0.040$ & $+5.012$ &
      $+2.907$ & $+3.196$ \\
      $10\,c_2({}^1S_0)$ & $-$ & $+3.711$ & $-6.492$ &
      $-2.781$ & $+1.377$ \\
      $10^2\,c_4({}^1S_0)$ & $-$ & $+2.276$ & $+3.523$ &
      $+5.798$ & $+0.524$ \\
      $10\,c_2({}^1P_1)$ & $-$ & $-5.604$ & $+10.526$ &
      $+4.922$ & $+0.996$ \\
      \hline \hline
      $c_0({}^3S_1)$ & $-1.040$ & $-0.545$ & $+4.521$ &
      $+2.936$ & $+3.748$ \\
      $10\,c_2({}^3S_1)$ & $-$ & $+2.115$ & $-5.150$ &
      $-3.035$ & $-3.091$ \\
      $10\,c_{t2}({}^3S_1)$ & $-$ & $+1.789$ & $+1.102$ &
      $+2.891$ & $+3.310$ \\
      \hline \hline
      $10\,c_2({}^3P)$ & $-$ & $+0.049$ & $-1.968$ &
      $-1.919$ & $-1.946$ \\
      $10^2\,c_{t2}({}^3P)$ & $+5.040$ & $-6.045$ & $-3.543$ &
      $-4.547$ & $-3.874$ \\
      $10\,c_{ls2}({}^3P)$ & $-$ & $+7.328$ & $+0.254$ &
      $+7.582$ & $+9.531$ \\
      $10^3\,c_{t4}({}^3P)$ & $-$ & $+1.976$ & $-9.275$ & 
      $-7.299$ & $+1.296$ \\
      \hline \hline
    \end{tabular}
  \end{center}
  \caption{Same as Table~\ref{tab:couplings-III} but for the modified
    Weinberg counting defined by Eqs.~(\ref{eq:delta-Vc-lo-III})
    and (\ref{eq:delta-Vc-nlo-III}). Being a local contact-range
    potential (plus a spin-orbit term, which is only local in r-space),
    the previous modifications affect the singlet and
    triplet partial waves simultaneously. Hence the two groups of
    couplings acting on the $\{ {}^1S_0, {}^1P_1 \}$ and
    $\{ {}^3S_1-{}^3D_1, {}^3P_0, {}^3P_1, {}^3P_2-{}^3F_2 \}$
    change with respect to Table~\ref{tab:couplings-III}.
    The finite-range part of the potential is not modified and identical
    to that of Ref.~\cite{Gezerlis:2014zia}.
  }
  \label{tab:couplings-III-mod}
\end{table}

\section{Error estimations}
\label{app:err}

In this Appendix I explain how to estimate errors in pionful EFT.
I begin with the observation that the quantity around which I organize
calculations is the perturbative expansion of
the cotangent of the phase shift.
The $\nu$-th term in this expansion
can be schematically written as
\begin{eqnarray}
  [\cot{\delta}]^{(\nu)} = 2 \mu \, k\,
  \sum_{{\substack{\nu_1, \nu_2 \geq 0 \\ \nu_1 + \nu_2 = \nu}}}\,
  \langle \Psi^{(0)} | V^{(\nu_1)} |  \Psi^{(\nu_2)} \rangle \, ,
  \nonumber \\
\end{eqnarray}
with $| \Psi^{(\nu)} \rangle$ the expansion of the wave function
\begin{eqnarray}
  | \Psi \rangle = \sum_{\nu \geq 0}\,| \Psi^{(\nu)} \rangle \, .
\end{eqnarray}
From this expression (plus a suitable choice of the normalization of the wave
function) one can reproduce the perturbative series of
the phase shifts in p- and r-space.
If one figures out how the terms in the expansion of the cotangent do scale,
the truncation errors are a trivial matter.

\subsection{Expected scaling properties}

One expects the scaling of the terms in the expansion of
the cotangent to be
\begin{eqnarray}
  [\cot{\delta}]^{(\nu)} = c\, {\left( \frac{Q}{M} \right)}^{\nu}
  \quad \mbox{with} \quad c = \mathcal{O}(1) \, , \label{eq:cotd-scaling}
\end{eqnarray}
with $Q$ representing the light scales, $M$ the breakdown scale and
$c$ a dimensionless number of order one.
This scaling is the result of considering the EFT expansion as applied to
a dimensionless quantity such as the cotangent of the phase shift.

Among the light scales represented by $Q$,
the most important ones are the momentum $k$ and the pion mass $m_{\pi}$.
However they do not necessarily have the same {\it weight}.
In particular I propose the substitution rule
\begin{eqnarray}
  \frac{Q}{M} \to \{ \frac{k}{M} , \frac{m_{\pi}}{2 M} \} \, ,
  \label{eq:Q-exp}
\end{eqnarray}
which implies that the EFT expansion in the pion mass is
more convergent than the expansion in the momentum.
The reason why I use different numerical factors for the expansions in $k$
and $m_{\pi}$ comes from the analysis of a generic two-potential model
that I present at the end of this Appendix.

\subsection{Truncation errors in the phase shifts}

The expansion of the phase shift and its cotangent are related by
\begin{eqnarray}
  \delta^{(\nu)} &=& - (\sin^2{\delta^{(0)}})\,
        [\cot{\delta}]^{(\nu)} + \delta_R^{(\nu)} \, ,
\end{eqnarray}
where $\delta_R^{(\nu)}$ indicates the terms that come from reexpanding
the lower order corrections.
These $\delta_R^{(\nu)}$ terms are not essential for the error estimations
as they are expected to show the same scaling as
$[\cot{\delta}]^{(\nu)}$.
Now, if one considers the EFT expansion of the phase shift
up to order $Q^{\nu_{\rm max}}$:
\begin{eqnarray}
  \delta_{\rm EFT} = \sum_{\nu=0}^{\nu_{\rm max}}\delta^{(\nu)} \pm 
  | \Delta\,\delta_{\rm EFT} | \, ,
\end{eqnarray}
the error in the phase shift can be estimated in several ways,
one of which is to exploit the expected scaling of
the cotangent
\begin{eqnarray}
  {\left( \Delta\,\delta_{\rm EFT} \right)}^{(I)}
  &=& (\sin^2{\delta^{(0)}})\,
  [\cot{\delta}]^{(\nu_{\rm max} + 1)} \, ,
  \nonumber \\
  &=& (\sin^2{\delta^{(0)}})\,
            {\left( \frac{Q}{M} \right)}^{\nu_{\rm max} + 1} \, , \label{eq:err-1}
\end{eqnarray}
in which the second line comes from the scaling of
$[\cot{\delta}]^{(\nu_{\rm max} + 1)}$ as written
in Eq.~(\ref{eq:cotd-scaling}),
where I set $|c| = 1$.
Another way is to use the lower order contribution to the phase shifts
to estimate the expected size of the missing terms.
For instance, if one uses the highest order computed
(i.e. $\nu = \nu_{\rm max}$), the truncation error is expected to be
\begin{eqnarray}
  {\left( \Delta\,\delta_{\rm EFT} \right)}^{(II)}
  = \left(\frac{Q}{M}\right)\,
  | \delta^{(\nu_{\rm max})} | \, . \label{eq:err-2}
\end{eqnarray}
With these two types of errors, one can estimate the truncation error to be
\begin{eqnarray}
  \Delta\,\delta_{\rm EFT} &=& {\rm max}
  \bigg[ 
    {\left( \Delta\,\delta_{\rm EFT} \right)}^{(I)} ,
    {\left( \Delta\,\delta_{\rm EFT} \right)}^{(II)}
    \bigg] \, , \label{eq:err-sum}
\end{eqnarray}
where ${\rm max}$ indicates the maximum of the two values.
For a well chosen $M$ both errors are expected to be of comparable size.
Finally,  the meaning of $Q/M$ depends on whether one expands in terms of
the external momentum or the pion mass.
For concreteness I will choose the following combination
\begin{eqnarray}
  {\left( \frac{Q}{M} \right)}^{\nu_{\rm max} + 1} \to
  \left[
    {\left( \frac{k}{M} \right)}^{\nu_{\rm max} + 1} +
    {\left( \frac{m_{\pi}}{ 2 M} \right)}^{\nu_{\rm max} + 1} \right]
    \, , \nonumber \\
\end{eqnarray}
in which the uncertainties associated to the momentum and pion mass
expansions are added coherently.
Of course, it is important to notice that this is not the only possible
method to estimate the errors of the EFT expansion
(see for instance Ref.~\cite{Epelbaum:2014efa}).
Yet I expect other alternative methods to yield comparable results.

Finally it is important to stress the statistical nature of
the error bands, which means that the way in which they are calculated
and the particular choice of $M$ are not necessarily uniquely defined.

\subsection{A two-potential model of the $Q/M$ expansion}
\label{subsec:two-potential}

Previously I have proposed a different convergence rate
for the pion mass and the momentum expansions.
This translates into two different interpretations of $Q/M$: $k/M$
and $m_{\pi}/ 2 M$, respectively.
To understand this distinction, I will analyze the $Q/M$ expansion
for a generic two-potential model of the type
\begin{eqnarray}
  V_F = V_L + V_S \, ,
\end{eqnarray}
where $V_L$ and $V_S$ are a long- and short-range potential which together
comprise the full potential $V_F$.
I will assume these two potentials to decay exponentially at
long distances, i.e.
\begin{eqnarray}
  V_L = f_L(m_L r)\,e^{-m_L r} \quad \mbox{,} \quad V_S = f_S(m_S r)\,e^{-m_S r}
  \, , \nonumber \\
\end{eqnarray}
with $m_L$ and $m_S$ the mass scale of $V_L$ and $V_S$, while $f_L$ and $f_S$
are functions which encapsulate all the other details of the potentials
with the exception of their ranges.

If one uses $\delta_F$, $\delta_S$ and $\delta_L$ to denote the phase shifts
generated by $V_F$, $V_S$ and $V_L$, one can understand the difference between
the full and the long-range phase shifts $\delta_F$ and $\delta_L$
as a matrix element of the short-range potential
\begin{eqnarray}
  \cot{\delta_F} - \cot{\delta_L} =
  2 \mu k\,\langle \Psi_F | V_S | \Psi_L \rangle \, ,
\end{eqnarray}
which is actually pretty similar to what one is calculating within EFTs.
In the expression above $| \Psi_F \rangle$ and $| \Psi_L \rangle$ are
the full and long-range wave functions generated by $V_F$ and $V_L$,
respectively.
The difference of the cotangents can be expanded as
\begin{eqnarray}
  \cot{\delta_F} - \cot{\delta_L}
  &=& \sum_{\nu} c^{(\nu)} \left( \frac{Q}{M} \right)^{\nu} \nonumber \\
  &=& \sum_{\nu_1, \nu_2} c^{(\nu_1, \nu_2)}
  \left( a_1 \frac{k}{m_S} \right)^{\nu_1} \,
  \left( a_2 \frac{m_L}{m_S} \right)^{\nu_2} \, , \nonumber \\
\end{eqnarray}
where in the second line I have explicitly indicated that
it is a power series with two expansion parameters.
The question is how to determine the radius of convergence of each of
the two series.

For answering this question one simply has to consider
the analytic properties of this matrix element:
\begin{eqnarray}
  \langle \Psi_F | V_S | \Psi_L \rangle \, .
\end{eqnarray}
For the external momentum $k$, one begins by noticing that at long distances
($m_L r \gg 1$) the wave functions behave as a superposition of
an incoming and outgoing spherical wave
\begin{eqnarray}
  \langle {r} | \Psi \rangle &\to&
  \langle {r} | \Phi \rangle =
  \sum_{\pm}\, \frac{e^{\pm i {k} {r}}}{k r}
  \, , 
\end{eqnarray}
where I use $| \Psi \rangle = | \Psi_F \rangle$, $| \Psi_L \rangle$
as generic notation for the full and long-range wave functions,
while $| \Phi \rangle$ stands for their asymptotic piece.
The previous implies that the infrared part of the integrand
in the matrix element will behave as
\begin{eqnarray}
  \langle \Psi | V_S | \Psi_L \rangle \sim
  \sum_{\pm} \int^{\infty}
  dr\, e^{-m_S r \, \pm \, 2 i k r} \, g(m_S r, Q r) \, , \nonumber \\
\end{eqnarray}
with $g$ a function containing details that are inessential
for the present analysis, and where only the upper integration
range is shown as I am only interested in the long range
properties of this integral.
The infrared convergence of this matrix element requires $| k | < m_S / 2$,
which coincides with what is expected from a modified
effective range formulation of two-body scattering.

The convergence with respect to the light mass $m_L$ requires
a bit more elaboration.
As in the previous case, the key factor to consider is
the asymptotic behavior of the wave functions.
While the longest-range piece of the wave functions is indeed
a superposition of an incoming and outgoing spherical wave,
the next-to-longest-range correction comes
from the range of the potential.
One can write
\begin{eqnarray}
  \langle r | \Psi \rangle =
  \langle r | \Phi \rangle + \langle r | \delta \Psi \rangle \, ,
\end{eqnarray}
with $| \Phi \rangle$ the asymptotic wave function and
$| \delta \Psi \rangle = (| \Psi \rangle - | \Phi \rangle)$
the difference between the full and asymptotic wave functions,
which in turn can be expanded as
\begin{eqnarray}
  \langle r | \delta \Psi \rangle =
  \langle r | G_0 V | \Phi \rangle + \mathcal{O}(V^2) \, ,
\end{eqnarray}
with $V$ the potential and $G_0$ the resolvent operator.
Owing to the finite range of the potential, $V = V_F$ or $V_L$,
the previous perturbative expansion also works as a range expansion,
where the behavior of the first range correction is
\begin{eqnarray}
   \langle r | G_0 V | \Phi \rangle \sim e^{-m_L r} \, (\dots) \, ,
\end{eqnarray}
for both $V_F$ and $V_L$ (with the {\it dots} representing other
non-exponential or shorter-range corrections), from which
it is possible to deduce the infrared behavior of the matrix elements.
Now I take into account the following decomposition
\begin{eqnarray}
  \langle \Psi_F | V_S | \Psi_L \rangle &=&
  \langle \Phi_F | V_S | \Phi_L \rangle 
  + \delta \, \langle \Psi_F | V_S | \Psi_L \rangle 
  + \mathcal{O}(\delta^2) \, , \nonumber \\
\end{eqnarray}
where I am interested in the term:
\begin{eqnarray}
  \delta \,  \langle \Psi_F | V_S | \Psi_L \rangle =
  \langle \delta \Psi_F | V_S | \Psi_L \rangle +
  \langle \Psi_F | V_S | \delta \Psi_L \rangle \, , \nonumber \\
\end{eqnarray}
which is the one that contains information about the $m_L$ expansion.
In fact, from its long-range behavior one finds that 
\begin{eqnarray}
  \delta \langle \Psi | V_S | \Psi_L \rangle \sim
  \int^{\infty} dr e^{-(m_L + m_S) r} \, h(m_S r, Q r) \, , \nonumber \\
\end{eqnarray}
with $h$ a function containing all the non-essential details.
This matrix element converges for $| m_L | < m_S$, which in hindsight is a
pretty obvious result: there can only be a light-mass expansion
if the light particle is indeed lighter than the heavy one.

Thus for this toy model the power counting expansion takes the form
\begin{eqnarray}
  \cot{\delta_F} - \cot{\delta_L}
  &=& \sum_{\nu_1, \nu_2} c^{(\nu_1, \nu_2)}
  \left( \frac{k}{m_S / 2} \right)^{\nu_1} \, \left( \frac{m_L}{m_S} \right)^{\nu_2} \, , \nonumber \\
\end{eqnarray}
which is equivalent to the following
\begin{eqnarray}
  \frac{Q}{M} \to \{ \frac{k}{m_S/2} , \frac{m_L}{m_S} \} \, .
\end{eqnarray}
Now if one makes the substitutions $m_L \to m_{\pi}$ and $m_S \to 2 M$,
one ends up with Eq.~(\ref{eq:Q-exp}).


%

\end{document}